\documentclass[12pt,a4paper]{article}
\usepackage{epsfig}
\usepackage{graphics}
\usepackage{subfigure}
\usepackage{a4wide}
\usepackage{amssymb}
\usepackage{amsmath}
\usepackage{amsfonts}
\makeatletter  

\@addtoreset{equation}{section}
\makeatother
\begin{document} 
\def\vecs{{\pmb{\sigma}}}
\def\vecS{{\pmb{\Sigma}}}
\def\vecalpha{{\pmb{\alpha}}}
\def\vecgamma{{\pmb{\gamma}}}
\def\vecvarepsilon{{\pmb{\varepsilon}}}
\def\vecnabla{{\pmb{\nabla}}}
\def\vecphi{{\pmb{\phi}}}
\def\vectau{{\pmb{\tau}}}
\def\vecomega{{\pmb{\omega}}}
\def\vecpi{{\pmb{\pi}}}
\def\vecPi{{\pmb{\Pi}}}
\begin{titlepage}
\vspace{2cm}
\title{
\vskip -50pt 
\rightline{\small{DAMTP-2000-54}}
\rightline{\small{hep-th/0006147}}
\vskip 40pt
\begin{bfseries}
$S^3$ Skyrmions and the Rational Map Ansatz
\end{bfseries}
}
\vspace{2cm}
\author{\\ S Krusch\footnote{email: S.Krusch@damtp.cam.ac.uk}\quad 
\bigskip\\
\textit{Department of Applied Mathematics and Theoretical Physics}\\ 
\textit{University of Cambridge}\\
\textit{Wilberforce Road, Cambridge CB3 0WA, England}\\
} 
\date{\large June 19, 2000}
\maketitle
\thispagestyle{empty}
\begin{abstract}
\noindent
This paper discusses multi-Skyrmions on the $3$-sphere $S^3$   
with variable radius $L$ using the rational map ansatz. For baryon
number $B = 3, \dots,9$ this 
ansatz produces the lowest energy solutions known so far.
By considering the geometry of the model we find an approximate
analytic formula 
for the shape function. This provides an insight why Skyrmions
have a shell-like structure.
\end{abstract}
\vspace{2cm}
\centerline{PACS-number: 12.39.Dc}
\vspace{4cm}
\begin{tt}
\pageref{lastref} pages, 14 figures
\end{tt}
\end{titlepage}

\section{Introduction}
\label{Introduction}

In this paper we discuss new developments in the $SU(2)$
Skyrme model \cite{Skyrme-61} and its generalization to the case where
the physical space is a $3$-sphere of radius $L$. 
The Skyrme model is a nonlinear field theory of mesons whose
field configurations are labelled by an integer, the topological
charge. This charge can be identified with the baryon number
$B$ \cite{Skyrme-61}. 
Static field configurations which minimize the energy of the Skyrme
model for a given baryon number $B$ are called
Skyrmions. When the theory is quantized, the Skyrme model not only
describes the proton and the neutron reasonably well \cite{Adkins-83,
Meier-97} but also larger nuclei \cite{Leese-95, Irwin-00}. 
However, in order to be able to perform the quantization it is
important to reach a good understanding of the classical
solutions. In \cite{Sutcliffe-97} Skyrmions were calculated
numerically for small baryon number $B$, and it was shown that they
have certain discrete symmetries. These symmetries have been confirmed
by the rational map ansatz \cite{Houghton-98} which also reproduces
the energies of the Skyrmions with good accuracy.

From a mathematical point of view, field configurations in the Skyrme
model are represented by maps from $({\mathbb R}^3~\cup~\{\infty\}) 
\cong S^3$ to $SU(2) \cong S^3$. 
Therefore, it is natural to generalize the model such that
physical space is a $3$-sphere of radius $L$. From a physical point of
view, a Skyrmion on a $3$-sphere describes a finite baryon density    
\cite{Manton-87}. Reducing $L$ increases the baryon
density. Varying $L$  the $S^3$ model exhibits phenomena such as
localization--delocalization
transitions and the restoration of
chiral symmetry. These results can be compared with Skyrme crystal
calculations \cite{Castillejo-89, Kugler-88, Kugler-89}, 
but are also interesting in their own right. 

In this paper we only consider static configurations and their
energies. In particular,
we generalize the rational map ansatz to describe
Skyrmions on $S^3$. The energies of our ansatz are the lowest ones
known so far.  
Moreover, these approximations have a well defined
limit for $L \to \infty$, namely, the rational map Skyrmions in flat
space. Geometric considerations lead to an analytic ansatz for the
shape function which is completely specified by a small set of
parameters. We  show that this ansatz agrees well with the numerical
solution and captures the behaviour of Skyrmions on $S^3$. 

In the following section we review the Skyrme model on
general $3$-dimensional manifolds. The first part focuses on the
geometrical meaning of the Skyrme energy. 
The next part clarifies the relationship between the geometric
formulation in flat space and the standard formulation in terms of the
Lie group $SU(2)$. In section \ref{RationalMaps} the model is
generalized to the $3$-sphere. We describe the rational map ansatz in
detail and also recall the doubly axially-symmetric ansatz 
\cite{Jackson-89}.
In section \ref{ConformalMap} we derive an analytic ansatz for the
shape function and discuss its symmetries. For special
values of $B$ the energy can be calculated explicitly as a function of
the radius $L$. 
In section \ref{NumericalResults} we calculate the shape function 
numerically. We also compare the numerical
shape function to the analytic shape function of the previous
section. Finally, we use the analytic shape function to approximate
Skyrmions in ${\mathbb R}^3$.
In section \ref{Phasetransitions} we discuss chiral symmetry and phase
transitions. 

\section{Geometry of Skyrmions}
\label{Geometry}

In this section we describe static solutions of the Skyrme model
on general $3$-dimensional manifolds. 
We follow Manton's approach \cite{Manton-87} and present a 
geometric point of view. 
First, we introduce the geometric notion of the
strain tensor and construct the Skyrme energy for general
manifolds. Then we discuss the properties of the energy density and
derive some formulae which will be important in the following
sections. 
Finally, we show that in flat space the geometric energy density is
equivalent to the standard energy density. 

A field configuration is a map $\pi$ from a physical space $S$
to a target space $\Sigma$. Both $S$ and $\Sigma$ are $3$-dimensional 
Riemannian manifolds which are connected and orientable, and their
metrics are $t_{ij}$ and $\tau_{\alpha \beta}$,
respectively. We denote by $x$ a point in $S$ and $\pi(x)$ its
image in $\Sigma$.
Choosing dreibeins ${e_m}^i(x)$ on $S$ and
${\zeta_\mu}^\alpha(\pi(x))$ on $\Sigma$, we can define the Jacobian
matrix 
\begin{equation}
J_{m \mu} (x) = {e_m}^i (x) (\partial_i \pi^\alpha(x)) \zeta_{\mu
\alpha}(\pi(x)).  
\end{equation}
The matrix $J_{m \mu} (x)$ is a measure of the deformation induced by
the map $\pi$ at the point $x$ in $S$.
However, it is not unique since a rotation of the dreibeins
${\zeta_\mu}^\alpha(\pi(x))$ does not affect the geometrical
deformation. This leads us to define the strain tensor  
\begin{equation}
\label{straintensor}
D_{m n} = (J J^T)_{m n} 
= {e_m}^i (\partial_i \pi^\alpha) {e_n}^j (\partial_j 
\pi^\beta) \tau_{\alpha \beta}
\end{equation}
which only depends on the metric on $\Sigma$. Here and in the
following we suppress the direct reference to $x$.
The strain tensor is symmetric and positive semi-definite\footnote{
Generically, $J$ is non-degenerate so that $D$ is
positive definite. However, there are submanifolds of zero baryon
density, i.e. $\det(J(x)) = 0$. For further discussion see
\cite{Houghton-00}.} 
but it is not invariant under rotation of the dreibeins
${e_m}^i$. In fact, under an orthogonal transformation $O$,
$D$ transforms into $O^T D O$. A well known result from linear algebra
is that the characteristic polynomial $P(\lambda) = \det(D - \lambda
I)$ is invariant under orthogonal transformations. Denoting the
non-negative eigenvalues of the strain tensor $D$ by $\lambda_1^2$,
$\lambda_2^2$, and $\lambda_3^2$ we obtain the following invariants of
$D$: 
\begin{equation}
\label{deftrace}
\begin{array}{l}
{\rm Tr}~ D = \lambda_1^2+\lambda_2^2+\lambda_3^2   \\ \\
\frac{1}{2}({\rm Tr}~ D)^2 - \frac{1}{2}{\rm Tr}~ D^2 =
\lambda_1^2\lambda_2^2+\lambda_2^2\lambda_3^2+\lambda_1^2\lambda_3^2
\\ \\
\det D = \lambda_1^2 \lambda_2^2 \lambda_3^2.
\end{array}
\end{equation}
As we will demonstrate in this section, choosing the energy of a field
configuration to be the integral of the following sum of these
invariants generalizes the Skyrme model to arbitrary manifolds $S$ and
$\Sigma$:
\begin{equation}
\label{Egeo}
E = \int_S \lambda_1^2+\lambda_2^2+\lambda_3^2
+ \lambda_1^2\lambda_2^2+\lambda_2^2\lambda_3^2+\lambda_1^2\lambda_3^2
\end{equation}
where the integration measure is $\sqrt{\det t}~ {\rm d}^3 x$. The first
three terms are only quadratic in the derivatives and correspond to
the ``sigma model term''. The configurations which are minimizing this
term are known as harmonic maps \cite{Eells-64}.
Geometrically, $\pi$ induces a linear map
$\pi^*$ mapping the unit vector ${e_m}^i$ which is formed by the
$m$th dreibein to the vector ${e_m}^i \partial_i \pi^\alpha$. 
The sum of the squared lengths of these vectors is the ``sigma model
term''. Thus, only the lengths of the vectors $\pi^*({e_m}^i)$ are
important.

The geometric meaning of the quartic terms can be understood by
considering the area element $\epsilon^{qmn} {e_m}^i {e_n}^j$ formed
by two dreibeins. This area element is mapped to the following 
area element in $\Sigma$:
\begin{equation}
\epsilon^{qmn} {e_m}^i {e_n}^j 
\partial_i \pi^\alpha \partial_j \pi^\beta.
\end{equation}
The sum of the squares of these area elements is proportional 
to the ``Skyrme term''. This term is also a measure of the angular
distortion of the map $\pi^*$.

So far only the first two invariants of the strain tensor have been
used. The third invariant is the square of the determinant
of the Jacobian, $(\det J)^2$. 
Since the manifolds $S$ and $\Sigma$ are orientable, $\det{J}$ is
globally well defined so that we can set $\sqrt{\det D} = \det J$. 
$\det J$ locally changes the integration
measure on $S$ into the measure on $\Sigma$. Therefore, the integral
of $\det J$ over $S$ is equal to the volume ${\rm Vol} (\Sigma)$ 
times the degree of the map $\pi$. 
The degree of a map is a topological invariant which, roughly
speaking, measures how many preimages a point $\pi(x)$ of $\Sigma$ has
in $S$, and hence takes integer values. 
Any field configuration can be
labelled by its degree. In the Skyrme model the degree of the map
$\pi$ is identified with the baryon number $B$, and we obtain
\begin{equation}
\label{Bgeo}
B= \frac{1}{{\rm Vol} (\Sigma)} \int_S \lambda_1 \lambda_2 \lambda_3.
\end{equation}
We will call classical field configurations which minimize
the energy (\ref{Egeo}) Skyrmions. It is worth pointing out that there
is an alternative expression for equation (\ref{Egeo}). By
``completing the square'' we can rearrange the $\lambda_i$ in the
following way:  
\begin{equation}
\label{Egeo2}
E = \int_S  (\lambda_1 \pm \lambda_2 \lambda_3)^2 +
               (\lambda_2 \pm \lambda_3 \lambda_1)^2 +
               (\lambda_3 \pm \lambda_1 \lambda_2)^2 \mp
              6 \lambda_1 \lambda_2 \lambda_3. 
\end{equation}
Either the upper or the lower signs are chosen such that the integral
over the last term in equation (\ref{Egeo2}) is non-negative. 
Applying equation (\ref{Bgeo}), this integral is just $6\ |B|\ {\rm
Vol} (\Sigma)$.   
Therefore, the energy is bounded below by the so-called Faddeev-Bogomolny
bound \cite{Faddeev-76}:
\begin{equation}
\label{Bogbound}
E \ge 6\ |B|\ {\rm Vol} (\Sigma).
\end{equation}
Equation (\ref{Egeo2}) also gives rise to two sets of three Bogomolny
equations which are satisfied if and only if the Faddeev-Bogomolny bound is
attained,
\begin{equation}
\lambda_1 = \mp \lambda_2 \lambda_3,~ \lambda_2 = \mp \lambda_3 \lambda_1~~
{\rm and}~~
\lambda_3 = \mp \lambda_1 \lambda_2.
\end{equation}
The only non-trivial solutions of both sets of equations 
are $\lambda_1^2 = \lambda_2^2 = \lambda_3^2 = 1$, and it follows
that $B = \lambda_1 \lambda_2 \lambda_3 = \mp 1$.
Therefore, the strain tensor is the identity map, and the map $\pi$ is
an isometry.  
For $\Sigma = SU(2)$, this case can only occur if the
physical space is a $3$-sphere of radius $L=1$. This has already been
proven in \cite{Manton-Ruback}. 

The relative importance of the ``sigma model term'' and the ``Skyrme
term'' can be seen from their scaling behaviour. For this purpose
we consider a family of metrics $L^2\ t_{ij}$ where $L$ is a constant
length scale.
The dreibeins ${e_m}^i$ are,  informally speaking, the square roots
of the inverse metric $\frac{1}{L^2}\ t^{ij}$ of $S$. 
Therefore, they are
proportional to $\frac{1}{L}$ so that the eigenvalues $\lambda_i^2$ of
the strain tensor $D_{i j}$ are proportional to $\frac{1}{L^2}$. 
Since the measure $\sqrt{\det(L^2\ t_{ij})}\ {\rm d}^3 x$ scales with
$L^3$, the ``sigma model term''  scales like $L$, whereas the Skyrme
term scales like $\frac{1}{L}$.  
For large radius $L$ one might expect the ``sigma model term'' to be
dominant. Yet, this is not the case because, as we shall see, the
Skyrmion becomes localized and --- just as in flat space --- both of
these terms are equally important. 
However, for small radius $L$ the configuration is delocalized, and the
``Skyrme term'' will become dominant.

In the following, we relate this geometric formulation to the
standard Skyrme model in flat space. 
For the remainder of this section $S$ will correspond to the flat
space ${\mathbb R}^3$ and $\Sigma$ to the Lie group $SU(2) \cong
S^3$. The basic field is the $SU(2)$ valued field 
\begin{equation}
\label{U}
U({\bf x}) = \sigma({\bf x}) + i \vecpi({\bf x}) \cdot \vectau
\end{equation}
where $\tau_i$ are the Pauli matrices which are Hermitian and  
satisfy the algebra 
$\tau_i \tau_j = \delta_{i j} + i \epsilon_{i j k} \tau_k$. Since $U$
is an element of $SU(2)$, the fields $\sigma$ and $\vecpi$ obey the
constraint $\sigma^2 + \vecpi^2 =1$. It is worth noting that 
$\vecpi$ in (\ref{U}) play the same role as the coordinates $\pi_i$ in the
discussion above, and $\sigma$ is just a function of those coordinates. 
The static solutions of the Skyrme model can be derived by varying the
following energy \cite{Houghton-98}:
\begin{equation}
\label{Estandard}
E = \int \left( -\frac{1}{2} {\rm Tr} \left( R_i R_i \right) - 
\frac{1}{16} {\rm Tr} \left( [R_i, R_j][R_i,R_j] \right) 
\right) {\rm d}^3 {\bf x}
\end{equation}
where $R_i = (\partial_i U) U^\dagger$ is an $su(2)$ valued current. 
A static solution of the variational equations could also be a saddle
point. Only solutions which minimize the energy are  called Skyrmions.
In order to show that the energies (\ref{Egeo}) and (\ref{Estandard})
are equivalent, we first relate the strain tensor $D_{i j}$ to the
current $R_i$. 
Since $\sigma$ is determined by $\vecpi$ we can calculate the induced 
metric $\tau_{\alpha \beta}$
on the target space $\Sigma$ in terms of the fields $\vecpi$:
\begin{equation}
\label{metrictau}
\tau_{\alpha \beta} = \delta_{\alpha \beta} + 
\frac{\pi_\alpha \pi_\beta}{\sigma^2}.
\end{equation}
Starting from equation (\ref{straintensor}) and noting that in flat
space the dreibeins can be chosen to be Kronecker deltas we obtain
the following expression: 
\begin{eqnarray}
\label{D1}
D_{i j} &=& \partial_i \pi^\alpha \partial_j \pi^\beta \tau_{\alpha
\beta} \\
\nonumber \\
\label{D2}
&=& \partial_i \sigma \partial_j \sigma + 
\partial_i \vecpi \cdot \partial_j \vecpi \\
\nonumber \\
\label{D3}
&=& \frac{1}{2} {\rm Tr} \left(
(\partial_i \sigma + i \vectau \cdot \partial_i \vecpi)
(\partial_j \sigma - i \vectau \cdot \partial_j \vecpi) 
\right) \\
\nonumber \\
\label{D4}
&=& \frac{1}{2} {\rm Tr} \left( 
\partial_i U \partial_j U^\dagger 
\right) \\
\nonumber \\
\label{D5}
&=& - \frac{1}{2} {\rm Tr} \left( R_i R_j \right).
\end{eqnarray}
Equation (\ref{D2}) follows from (\ref{D1}) by using the chain rule
and the formula for the metric (\ref{metrictau}). 
Conceptually, the step from equation (\ref{D2}) to equation (\ref{D3})
is very important. The metric $\tau_{\alpha \beta}$ 
is expressed with the group
multiplication and the trace which provides a scalar product in
$su(2)$. This is the transition from geometric to Lie group language.
Equation (\ref{D4}) is a trivial
consequence of definition (\ref{U}), and the last equation follows from
$U U^\dagger = 1$.
Now, we can show that the two energy densities are equal. 
Since $R_i$ is an $su(2)$ current, and therefore
anti-Hermitian, it can be expressed as
$R_i = R_i^\alpha~ i \tau_\alpha$.
We start with the energy density in (\ref{Estandard}) 
and expand $R_i$ in terms of
Pauli matrices:
\begin{eqnarray}
\label{Ed1}
 -\frac{1}{2} {\rm Tr} \left(R_i^\mu R_i^\nu 
i \tau_\mu i \tau_\nu \right) - 
\frac{1}{16} {\rm Tr}
\left(R_i^\mu R_j^\nu 
[i \tau_\mu, i \tau_\nu]
R_i^{\mu^\prime} R_j^{\nu^\prime} 
[i \tau_{\mu^\prime},i \tau_{\nu^\prime}]
\right) \\
\nonumber \\
\label{Ed2}
= R_i^\mu R_i^\mu + \frac{1}{4} {\rm Tr}
\left(R_i^\mu R_j^\nu 
\epsilon_{\mu \nu \rho} \tau_\rho
R_i^{\mu^\prime} R_j^{\nu^\prime} 
\epsilon_{\mu^\prime \nu^\prime \rho^\prime} \tau_{\rho^\prime} 
\right) \\
\nonumber \\
\label{Ed3}
= D_{i i} + \frac{1}{2} \left(R_i^\mu R_i^\mu
R_j^\nu R_j^\nu
- R_i^\mu R_j^\mu R_i^\nu R_j^\nu \right) \\
\nonumber \\
\label{Ed4}
= D_{i i} + \frac{1}{2} \left(D_{ii} D_{jj} - D_{ij} D_{ji} 
\right) \\
\nonumber \\
\label{Ed5}
= {\rm Tr}~ D + \frac{1}{2} \left( {\rm Tr}~ D \right)^2 
- \frac{1}{2} {\rm Tr}~ D^2. 
\end{eqnarray}
Equation (\ref{Ed2}) follows from (\ref{Ed1}) by inserting the
commutation relations of the Pauli matrices $\vectau$. In the
following equation we use the contraction
$\epsilon_{ijk} \epsilon_{ilm}= \delta_{jl}
\delta_{km}-\delta_{jm}\delta_{kl}$. 
In equation (\ref{Ed4}) we apply the result (\ref{D5}). The
last equality follows from the definition of the trace. Equation
(\ref{Ed5}) is the energy (\ref{Egeo}) because of (\ref{deftrace}). 
Therefore, we have shown the equivalence of the two approaches for $S=
{\mathbb R}^3$. 
It is worth noting
that in the geometric picture the ``Skyrme term'' is related to the
square of an area element, whereas in the standard approach it
depends on the structure constants of the Lie algebra
$su(2)$. Both interpretations lead to generalizations. While in our
geometric interpretation it is natural to consider different
$3$-dimensional manifolds for $S$, and maybe for $\Sigma$, the
Lie algebra approach leads to generalizing the Lie group $SU(2)$ 
for example to $SU(N)$.       

\section{Skyrmions on $S^3$ and Rational Maps}
\label{RationalMaps}

In the previous section we have established the equivalence of the two
energies (\ref{Egeo}) and (\ref{Estandard}) in ${\mathbb R}^3$. 
In this section we will use (\ref{Egeo}) to generalize the model to
$S=S^3_L$ such that from now on physical space is a 3-sphere of radius
$L$. This includes the original model if we take the limit $L \to
\infty$. We also fix $\Sigma = SU(2) \cong S^3_1$.

The Skyrme model cannot be solved analytically either in flat space,
or on $S^3_L$, apart from the case if $L=1$ and $B=1$ as mentioned above. 
However, in flat space there
are analytic ans\"atze which give good qualitative and quantitative
agreement with exact solutions obtained numerically. 
By ansatz we mean a test function that minimizes the
energy within a given class of test functions. 
The lower the energy the better we
expect the ansatz to approximate the exact solution.
In the following we will generalize the rational map
ansatz \cite{Houghton-98}, which has been very successful in flat
space, to $S^3_L$. 

A rational map is a holomorphic function from $S^2 \to S^2$. Treating
each $S^2$ as a Riemann sphere, with complex coordinates $z$ and $R$,
respectively, a rational map can be written as
\begin{equation}
R(z) = \frac{p(z)}{q(z)}
\end{equation}
where $p(z)$ and $q(z)$ are polynomials in $z$ which are assumed to
have no common factors. It has been shown by Donaldson
\cite{Donaldson-84}, and also by Jarvis \cite{Jarvis-96}, that there is a
one-to-one correspondence between rational maps and
monopoles. In flat space it has been found that many solutions of the
Skyrme equation with baryon number $B$ look rather like monopoles with
monopole number equal to $B$. Therefore, rational maps can be used to
approximate Skyrmions.

A point $x$ on $S^3_L$ is labelled by polar coordinates
$(\mu,\theta,\phi)$ such that
\begin{eqnarray}
x &=& \left(L \sin \mu \sin \theta \cos \phi, 
     L \sin \mu \sin \theta \sin \phi,
     L \sin \mu \cos \theta, 
     L \cos \mu \right) \\
\nonumber \\
 &=& \left(L \sin \mu~ {\hat {\bf n}}(\theta,\phi), L \cos \mu
\right)
\end{eqnarray}
where $\mu$, $\theta \in [0,\pi]$, and $\phi \in [0, 2 \pi]$ and ${\hat
{\bf n}}(\theta,\phi)$ is the unit vector on $S^2$. 
The $3$-sphere can be thought of as a collection of $2$-spheres
with varying radius equal to $L \sin \mu$.
With the stereographic projection $z = \tan \frac{\theta}{2} {\rm
e}^{i \phi}$ the $S^2$ can be identified with a Riemann sphere using a
single complex coordinate $z$. Alternatively, we can express the unit
vector ${\hat {\bf n}}$ in terms of $z$:
\begin{equation}
\label{normalv}
{\hat  {\bf n}(z)} = \frac{1}{1 + |z|^2}
 \left(
2 \mbox{Re} (z), 2  \mbox{Im} (z), 1 - |z|^2
\right).
\end{equation}
Similarly, points in the target $S^3$ can be labelled by $(f,R)$ where
$f$ is an 
angular variable analogous to $\mu$, 
and $R$ is a complex coordinate. The rational map
ansatz simply states that $f=f(\mu)$ and $R=R(z)$. 
This is only
consistent if $\sin f(\mu)$ vanishes where $\sin \mu$ does, {\it i.e.} $f(0)
= N_1 \pi$ and $f(\pi) = N_2 \pi$. The integer $N_f = N_1-N_2$ is a
topological invariant and cannot be changed by deforming $f(\mu)$
smoothly. 
In order to have a good limit for
$L \to \infty$ we fix $f(\mu)$ such that $N_2 = 0$ and set $N_1 = N_f$. 
In analogy to flat space we define a Skyrmion to have $N_f>0$,
whereas for an anti-Skyrmion $N_f < 0$. Since Skyrmions and anti-Skyrmions 
are related by reflection we will only consider Skyrmions
from now on. Therefore, our complete boundary conditions are:
\begin{eqnarray}
\nonumber
f(0) &=& N_f \pi ~~~ {\rm where} ~~~ N_f > 0 \\
\nonumber \\
\label{bc}
f(\pi) &=& 0.
\end{eqnarray}
In contrast to the flat case these boundary conditions do not follow
from a regularity argument. They are an artifact of our ansatz and
have to be handled with care (see \cite{Loss-87} for a discussion).

If we now use the notation of equation (\ref{U}) we can write the Skyrme
field in the following way:
\begin{eqnarray}
\label{Urat1}
U(\mu,z,{\bar z}) &=& \cos f(\mu) + i\ \sin f(\mu)\  
{\hat {\bf n}}(R(z)) \cdot \vectau \\
\nonumber \\
\label{Urat2}
&=& \exp
\left(
i\ f(\mu)~ {\hat {\bf n}}(R(z)) \cdot \vectau
\right) 
\end{eqnarray}
where ${\hat {\bf n}}(R(z))$ is as in equation (\ref{normalv}).
Equation (\ref{Urat2}) looks quite similar to the well known
spherically symmetric hedgehog ansatz \cite{Skyrme-61}. 
In fact, the hedgehog ansatz corresponds to the special case
$R(z) = z$.

Now, we can apply the formulae of the previous section.
The rational map ansatz gives rise to the following eigenvalues
$\lambda_i^2$ of the strain tensor (\ref{straintensor}): 
\begin{equation}
\lambda_1 = - \frac{f^\prime}{L}~~~{\rm and}~~~
\lambda_2 = \lambda_3 = 
\frac{\sin f}{L \sin \mu} \frac{1 + |z|^2}{1 + |R|^2} 
\left| \frac{{\rm d} R}{{\rm d} z} \right|.
\end{equation}
The minus sign in the expression for $\lambda_1$ is a consequence of
our definition of positive baryon number in (\ref{bc}).
One advantage of the rational map ansatz is the
decoupling of the radial and the angular strains.
Starting from formula (\ref{Bgeo}) the baryon number $B$ can be
written as a product of two integrals, one over $\mu$ and one over $z$
and ${\bar z}$ in the following way:
\begin{eqnarray}
\label{Brational}
B &=& \frac{2}{\pi} \int \left(
   -f^{\prime} \sin^2 f \right) {\rm d}\mu~ 
   \frac{1}{4 \pi} \int \left(
    \frac{(1+|z|^2)}{(1+|R|^2)}\left| \frac{{\rm d} R}{{\rm d} z} \right|
    \right)^2 \frac{2i~{\rm d}z {\rm d}{\bar z}}{(1+ |z|^2)^2}  \\
\nonumber \\
\label{B=NfNR}
  &=& N_f N_R.
\end{eqnarray}
The integral over $\mu$ is the integer $N_f$.  
The second integral is the pull-back of the area form on the
target sphere of the rational map $R(z)$. Taking the normalization 
into account this is just the degree $N_R$ of the rational map
$R(z)$. In fact, it can be shown that if $p(z)$ has the degree $n_p$
and $q(z)$ has the degree $n_q$ then $N_R = {\rm max}(n_p,n_q)$.

The energy can now be obtained from formula (\ref{Egeo}):
\begin{eqnarray}
\label{Erational0}
E = \int \left[
\frac{f^{\prime 2}}{L^2} \right.
 &+& 2 \left(\frac{f^{\prime 2}}{L^2} + 1
\right)
\frac{\sin^2 f}{L^2 \sin^2 \mu} 
\left(
\frac{1+ |z|^2}{1+|R|^2} \left| \frac{{\rm d} R}{{\rm d} z} \right|
\right)^2  \\ 
\nonumber \\
\nonumber \\        
\nonumber
&+& \left.
\frac{\sin^4 f}{L^4 \sin^4 \mu} 
\left(
\frac{1+ |z|^2}{1+|R|^2} \left| \frac{{\rm d} R}{{\rm d}z} \right|
\right)^4
\right]
\frac{2i~{\rm d} z {\rm d} {\bar z}}
{(1+ |z|^2)^2} L^3 \sin^2 \mu~{\rm d}\mu.
\end{eqnarray}
Similarly to the baryon density, the integration over $z$ and ${\bar
z}$ and the integration over $\mu$ factorizes. Therefore, formula
(\ref{Erational0}) can be rewritten as 
\begin{equation}
\label{Erational}
E = 4 \pi \int 
\left(
f^{\prime 2} L \sin^2 \mu  
 + 2 N_R (\frac{f^{\prime 2}}{L} + L) \sin^2 f +
{\cal I} \frac{\sin^4 f}{L \sin^2 \mu} 
\right)
{\rm d} \mu
\end{equation}
where $N_R$ is the degree of the rational map, and ${\cal I}$ is 
the following special function on the space of rational maps:
\begin{equation}
\label{calI}
{\cal I}  = \frac{1}{4 \pi} \int \left(
\frac{1 + |z|^2}{1 + |R|^2} \left|\frac{{\rm d} R}{{\rm d} z} \right|
\right)^4
\frac{2i~{\rm d} z {\rm d}{\bar z}}{(1 + |z|^2)^2}.
\end{equation}
To minimize $E$ for a given baryon number  $B = N_f N_R$, 
one should first minimize ${\cal I}$ over the space of rational maps
of degree $N_R$. This calculation was performed in \cite{Houghton-98},
and the result is displayed in table \ref{tableA1} in the appendix.
Then the profile function $f(\mu)$ is found by
solving the following Euler-Lagrange equation:
\begin{eqnarray}
\label{Euler}
\nonumber
f^{\prime \prime}
\left(
\frac{2 N_R}{L^2}\sin^2 f + \sin^2 \mu
\right) &+&
2 f^\prime \sin \mu \cos \mu 
+ \frac{2 N_R}{L^2} {f^\prime}^2 \sin f \cos f  \\
\nonumber \\
&-& 2 N_R \sin f \cos f  - \frac{2 {\cal I} \sin^3 f \cos f}{L^2
\sin^2 \mu} = 0
\end{eqnarray}
where ${\cal I}$ now takes the constant value in table \ref{tableA1}.
We require $f(\mu)$ to be a solution of
(\ref{Euler}) non-singular in $[0,\pi]$.
Equation (\ref{Euler}) has to be solved numerically. It has regular
singular points at the end points, {\it i.e.} close to these points the
solution has the following power law behaviour:  
$f(\mu) \approx N_f \pi - A_\pm 
\mu^{\rho_\pm}$ for $\mu \approx 0$ and $f(\mu) \approx B_\pm
(\pi-\mu)^{\rho_\pm}$ for $\mu \approx \pi$. 
Here $\rho_\pm = \frac{1}{2}(\pm \sqrt{1+8 N_R}-1)$ and $A_\pm$
and $B_\pm$ are arbitrary constants. The solution $f(\mu)$ is regular
if the exponent is equal to $\rho_+$ at both end points.

Equation (\ref{Erational}) has an important discrete
symmetry. The transformation 
\begin{equation}
\label{fsymmetry}
f(\mu) \to N_f \pi - f(\pi - \mu) 
\end{equation}
transforms a solution of (\ref{Euler})
into a solution which is also compatible with
the boundary conditions (\ref{bc}). 
Geometrically, $\mu \to \pi -\mu$ is a
reflection at the plane through the equator in  physical space,
whereas  $f \to N_f \pi - f$ is a reflection in target space. 
This means that a solution
which is localized for example at the south pole $\mu = \pi$ is transformed to a
solution which is localized at the north pole $\mu =0$.
Therefore, for fixed $B$ there are two degenerate solutions unless the
transformation (\ref{fsymmetry}) is a symmetry of $f(\mu)$ in which
case there is only one symmetric solution. 
The symmetry (\ref{fsymmetry})  does not have an analogue in flat space. 
Also note that the transformations $f(\mu) \to f(\pi - \mu)$ and $f(\mu)
\to - f(\mu)$ take a Skyrmion with baryon number $B$ into an 
anti-Skyrmion with baryon number $-B$. 

When we derived the energy of the rational map ansatz we used equation
(\ref{Egeo}). This could be transformed into equation (\ref{Egeo2}) by
``completing the square''. Having now calculated the integral over
the rational map, we can complete the square in a different way and
re-express (\ref{Erational}) as 
\begin{eqnarray}
\nonumber
E = 4 \pi L \int_0^\pi 
\left[
\left(f^\prime \sin \mu + \frac{\sqrt{{\cal
I}}\sin^2 f}{L \sin \mu}\right)^2 + 2N_R \sin^2 f
\left(\frac{f^\prime}{L} +1 \right)^2 
\right] {\rm d } \mu \\
\nonumber \\
\label{Erat}
 - 8 \pi(2 N_R+\sqrt{{\cal I}}) \int_0^\pi  f^\prime \sin^2 f~ 
{\rm d} \mu. 
\end{eqnarray}
The second integral in (\ref{Erat}) can be evaluated using $f(0)
= N_f \pi$ and $f(\pi) = 0$. Since the first integral in (\ref{Erat})
is positive the second integral provides us with the energy bound
\begin{eqnarray}
\label{Bograt}
E &\ge& 4 \pi^2 (\sqrt{{\cal I}} + 2 N_R) N_f \\
\nonumber \\
  &\ge& 12 \pi^2 N_R N_f.
\end{eqnarray}
The last inequality is valid because ${\cal I} \ge {N_R}^2$ which can
easily be shown by applying the Schwartz inequality.
Therefore, inequality (\ref{Bograt}) ``improves'' the Faddeev-Bogomolny
bound (\ref{Bogbound}) where ${\rm Vol} (\Sigma) = 2 \pi^2$ and $B =
N_R N_f$.  
This rational map bound (\ref{Bograt}) is valid, if the fields obey
the rational map ansatz. 
Exact solutions satisfy neither the rational map ansatz nor
necessarily the bound (\ref{Bograt}). We will discuss this
bound further in section \ref{NumericalResults}.

Before ending this section we review one further ansatz.
The symmetry group of $S^3_L$, which is $O(4)$, 
contains an  $O(2) \times O(2)$ subgroup. 
Jackson {\it et al} used this symmetry to
obtain doubly axially-symmetric ans\"atze for Skyrmions of various
baryon numbers $B$ \cite{Jackson-89}. 
In order to make best use of the symmetry it is convenient to
parameterize the $3$-sphere with a different set of angles
$(\chi,\alpha,\beta)$. With these coordinates a point on $S^3_L$ can
be written as 
\begin{equation}
x= (L \sin \chi \cos \alpha, L \sin \chi
\sin \alpha, L \cos \chi \cos \beta, L \cos \chi \sin \beta)
\end{equation}
where $\chi \in [0,\frac{\pi}{2} ]$ and $\alpha, \beta \in
[0,2 \pi]$.
Now we can write an $O(2) \times O(2)$ symmetric field configuration
$U \in SU(2)$ as
\begin{equation}
\label{O(2)ansatz}
U = \sin g  \cos p \alpha + i \tau_3 \sin g  \sin p \alpha 
+ i\tau_1 \cos g  \cos q \beta + i\tau_2 \cos g  \sin q \beta
\end{equation}
where $p$ and $q$ are integers and $g = g(\chi)$ is a shape function. In
this ansatz the eigenvalues $\lambda_i$ of the strain tensor
(\ref{straintensor}) are 
\begin{equation}
\lambda_1 = \frac{g^\prime}{L} , ~~~ 
\lambda_2 = \frac{p \sin g}{L \sin \chi} ~~~ {\rm and} ~~~
\lambda_3 = \frac{q \cos g}{L \cos \chi}.
\end{equation}
Using equation (\ref{Egeo}) we obtain the energy 
\begin{eqnarray}
\label{EO(2)}
E &=& 2 \pi^2 L \int 
\left({g^\prime}^2 + \frac{p^2 \sin^2 g}{\sin^2 \chi} + \frac{q^2
\cos^2 g}{\cos^2 \chi} \right) \sin (2 \chi) {\rm d} \chi \\
\nonumber \\
\nonumber
&+& \frac{2 \pi^2}{L} \int
\left({g^\prime}^2 \left(\frac{p^2 \sin^2 g}{\sin^2 \chi} + \frac{q^2
\cos^2 g}{\cos^2 \chi} \right) + \frac{p^2 q^2 \sin^2 g \cos^2 g}{\sin^2
\chi \cos^2 \chi}\right) \sin (2 \chi) {\rm d} \chi
\end{eqnarray}
and the baryon number 
\begin{equation}
B = p q \int 2 g^\prime  \sin g \cos g~ {\rm d} \chi.
\end{equation}
The shape function $g(\chi)$ can be calculated numerically by
minimizing the energy $E$ subject to the boundary conditions $g(0) = 0$
and $g(\frac{\pi}{2}) = \frac{\pi}{2}$ for various $p$, $q$ and
$L$. Note that these boundary conditions for $g(\chi)$ imply that 
$B = pq$.  
In \cite{Jackson-89} the energy for a given baryon number $B$ was also
minimized with respect to the radius $L$. 
Some of the results are displayed 
in figure \ref{fig5} of section  \ref{NumericalResults}.
It was shown that the solutions are only stable as long as $L$ is
smaller than a $B$-dependent critical length $L_{crit.}$.  
Moreover, for large baryon number $B$ the minimal energy for the
optimal radius $L$ is larger
than the energy of $B$ well separated Skyrmions in ${\mathbb R}^3$. 
Therefore, in this situation the ansatz fails badly.
However, for small baryon number $B$ and small radius $L$ the ansatz
is quite successful. 
For $B=1$ it agrees with the known $O(4)$ symmetric solution.  
For $B=2$ it predicts an $O(2) \times O(2)$ symmetric solution with a
very low energy. Given that the exact $B=2$ solution in flat space possesses
an $O(2) \times {\mathbb Z}_2$ 
symmetry this configuration is likely to be the exact solution on $S^3$. 

One particular feature of configurations of the form
(\ref{O(2)ansatz}) is that the following order parameter $O_1$
vanishes: 
\begin{equation}
\label{order}
O_1 = \langle \sigma \rangle^2 + \langle \vecpi \rangle^2 = 0 
\end{equation}
where $\langle~ \rangle$ means the average over the physical $S^3_L$.
It also turns out that configurations with $p=q$ attain their minimal
energy at particularly small values of $L$. Moreover, for $p=q$ the
solution has a symmetry similar to (\ref{fsymmetry}). 
The transformation
\begin{equation}
\label{gsymmetry}
g(\chi) \to \frac{\pi}{2} - g(\frac{\pi}{2} - \chi)
\end{equation}
transforms solutions of the Euler-Lagrange equation for the energy
(\ref{EO(2)}) into each other. We will need these properties in
section \ref{Phasetransitions}.

\section{The Shape Function as a Quasi-Conformal Map}
\label{ConformalMap}

We discuss next the shape function $f(\mu)$ of the rational map
ansatz. Since solving equation (\ref{Euler}) numerically provides
little insight we derive an analytic shape function. This ansatz
approximates the numerical shape function fairly accurately and also
confirms geometric ideas. 

In \cite{Manton-87} Manton approximated the shape function of the
$B=1$ Skyrmion by a conformal map. With this ansatz he showed that a
delocalized $B=1$ Skyrmion on a $3$-sphere is unstable for
$L>\sqrt{2}$. In this section we will generalize this idea for higher
baryon number $B$.
Using the rational map ansatz we will
derive an ansatz for the shape function which is conformal in an
average sense and which we will call quasi-conformal. In the
following section, we will show that this quasi-conformal ansatz is a
good approximation to the numerically calculated shape function.

A Skyrmion is a map from physical space $S^3_L$ labelled by 
$(\mu, z)$ to a target $S^3_1 \cong SU(2)$ labelled by $(f,R)$.
A map between two $3$-spheres is conformal if their metrics only
differ by a conformal factor:
\begin{equation}
\label{Econformal}
L^2\ {{\rm d} \mu}^2 + L^2\ \sin^2 \mu 
\frac{2i~{\rm d}z~{\rm d}{\bar z}}{(1+z {\bar z})^2}
= \Omega (\mu,z,{\bar z})^2 \left(
{{\rm d} f}^2 + \sin^2 f 
\frac{2i~{\rm d}R~{\rm d}{\bar R}}{(1+R {\bar R})^2} 
\right).
\end{equation}
We are interested in fields which obey the rational map
ansatz, {\it i.e.} $f=f(\mu)$ and $R = R(z)$. 
Therefore, we make the following approximations to equation
(\ref{Econformal}).
Firstly, since $f(\mu)$ is a function of $\mu$ only, 
we consider a conformal factor $\Omega$
which is also only a function of $\mu$. Secondly, we recall that
according to equation (\ref{Brational}) the
integral over the target $S^2$ is just $4 \pi$ times the degree
$N_R$ of the rational map. 
Therefore, we replace the metric on the target $S^2$ by
$N_R$ times the metric on the physical $S^2$. Locally, this is not a
very good approximation, but averaged over the whole $S^2$ this
reproduces the correct result. Since $f(\mu)$ is independent of $z$
and ${\bar z}$ it can only detect the averaged value. With these
approximations equation (\ref{Econformal}) can be simplified:
\begin{equation}
\label{quasiconf}
L^2\ {{\rm d} \mu}^2 + L^2\ \sin^2 \mu 
\frac{2i~{\rm d}z~{\rm d}{\bar z}}{(1+z {\bar z})^2}
= \Omega (\mu)^2 \left(
{{\rm d} f}^2 + \sin^2 f\ N_R 
\frac{2i~{\rm d}z~{\rm d}{\bar z}}{(1+z {\bar z})^2} 
\right).
\end{equation}
Eliminating $\Omega(\mu)$ from equation (\ref{quasiconf}) we obtain the 
following differential equation for $f(\mu)$:
\begin{equation}
\label{confeq}
\left(f^\prime(\mu)\right)^2 \sin^2 \mu = N_R \sin^2 f(\mu).
\end{equation}
We are interested in solutions which obey the boundary conditions
(\ref{bc}). It is convenient to replace $\sin f(\mu)$ by 
$\sin (\pi - f(\mu))$ before taking the square root:
\begin{equation}
\label{confeq2}
f^\prime(\mu) \sin \mu = \pm \sqrt{N_R} \sin (\pi - f(\mu)).
\end{equation}
For the  negative sign the solution of equation (\ref{confeq2})
diverges at the boundary and can therefore be discarded.
Equation (\ref{confeq2}) is solved by separation of variables,
and we obtain
\begin{equation}
\label{confansatz}
f(\mu) = \pi - 2 \arctan \left(k \left( \tan \frac{\mu}{2} \right)^{\sqrt{N_R}} 
\right)
\end{equation}
where $k$ is a positive integration constant. A negative $k$ would lead to
a negative baryon number and is incompatible with the boundary
conditions (\ref{bc}) whereas $k=0$ is just the trivial solution with baryon
number $B=0$. The quasi-conformal shape function (\ref{confansatz})
satisfies our boundary conditions (\ref{bc}) if and only if  $N_f =1$.
With equation (\ref{B=NfNR}) we obtain $B = N_R$.
The shape function (\ref{confansatz}) has the following important property.
Using the identity
\begin{equation}
\label{arctanidentity}
\arctan{x} + \arctan \frac{1}{x} = \frac{\pi}{2}~~~{\rm for}~~~x \ge 0
\end{equation}
it is easy to show that  for $k=1$ the shape function
(\ref{confansatz}) is invariant under the discrete symmetry
(\ref{fsymmetry}): 
\begin{equation}
\label{fsym}
f(\mu) = \pi - f(\pi - \mu).
\end{equation}
This means that for $k=1$ the solution is neither localized at the
south pole nor at the north pole. For $N_R=1$ the solution is
delocalized over the whole sphere whereas for $N_R>1$ it is rather
localized at the equator $\mu = \frac{\pi}{2}$.
By Taylor expanding equation (\ref{confansatz}) near  $\mu=0$ we find
that the shape function $f(\mu)$ behaves like $f(\mu) \approx \pi - 2
k (\frac{\mu}{2})^{\sqrt{N_R}}$.     
Recall that for the exact solution of the Euler-Lagrange equation
(\ref{Euler}) we obtained $f(\mu) \approx \pi - A_+ \mu^{\rho_+}$ 
where $\rho_+ = -\frac{1}{2} + \frac{1}{2} \sqrt{1 + 8 N_R}$.
So at $\mu = 0$ the shape function (\ref{confansatz}) and the exact
solution have a similar behaviour. The same is true for the other
boundary $\mu = \pi$.

Now, we can substitute the shape function (\ref{confansatz}) into
the equation for the energy (\ref{Erational}) and obtain 
\begin{equation}
\label{Eansatz}
E = 16 \pi \left(
 3 L N_R I_1(k) 
+ \frac{4}{L} \left(2 {N_R}^2  + {\cal I} \right) I_2(k)
\right)
\end{equation}
where
\begin{equation}
\label{eqI1}
I_1(k) = \int_0^\infty \frac{2 k^2 y^{2 \sqrt{N_R}}}
{(1+k^2 y^{2 \sqrt{N_R}})^2(1+y^2)} 
{\rm d} y
\end{equation}
and
\begin{equation}
\label{eqI2}
I_2(k) =\int_0^\infty \frac{k^4 y^{4 \sqrt{N_R}-2}(1+y^2)}
{2 (1+k^2 y^{2 \sqrt{N_R}})^4} 
{\rm d} y.
\end{equation}
In (\ref{eqI1}) and (\ref{eqI2}) we have simplified the integrand by
substituting $y = \tan \frac{\mu}{2}$.  
The integrals $I_1(k)$ and $I_2(k)$ can be evaluated in closed form if
$\sqrt{N_R}$ is an integer.  
Therefore, we will concentrate on $N_R=1$, $4$ and $9$  in the
remainder of this section. 

\newpage

\begin{figure}[!hb]
\begin{center}
\subfigure[$f(\mu)$ for $B=1$]{
\includegraphics[height=70mm,angle=270]{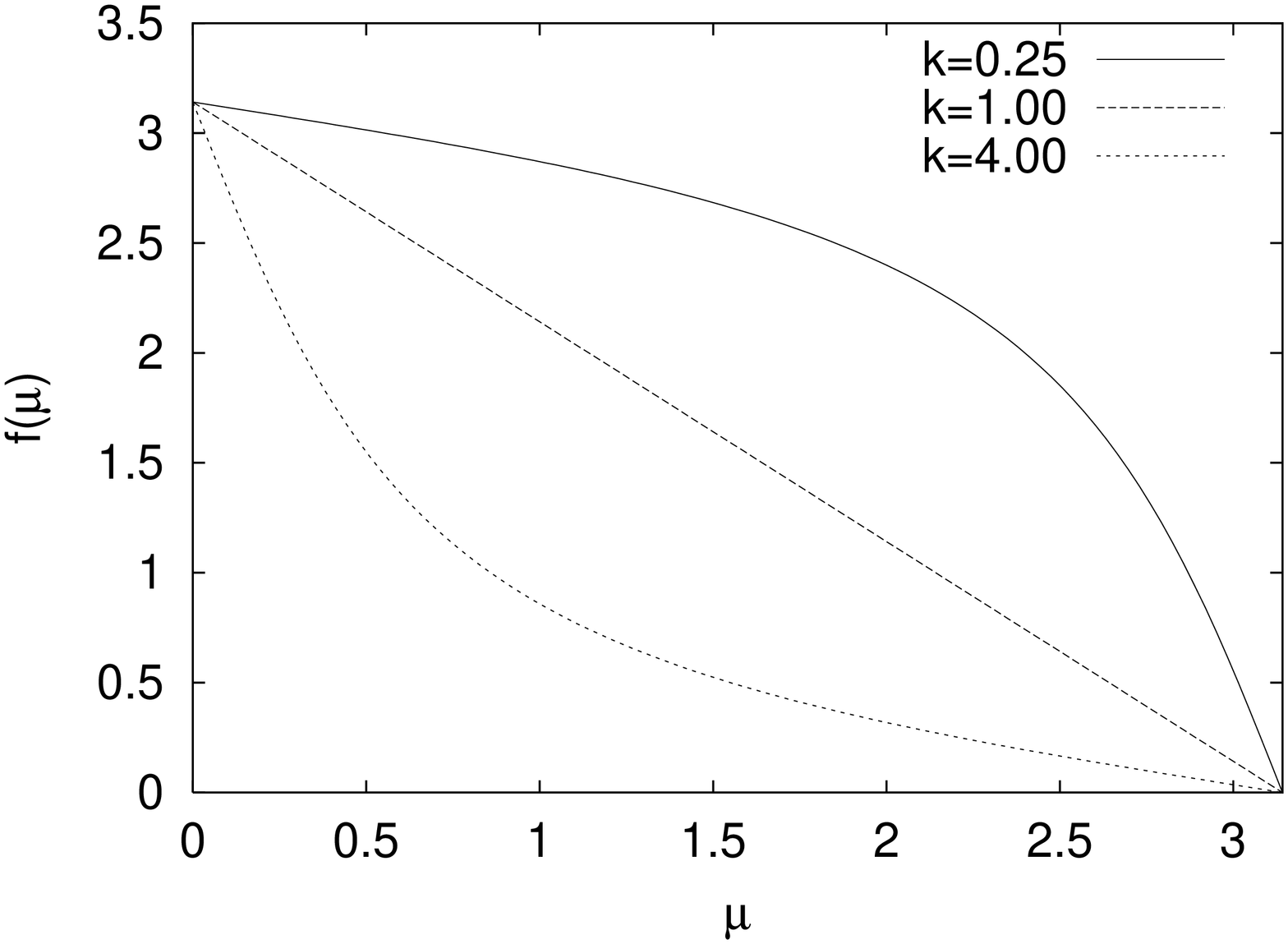}
}
\quad
\subfigure[$f(\mu)$ for $B=4$]{
\includegraphics[height=70mm,angle=270]{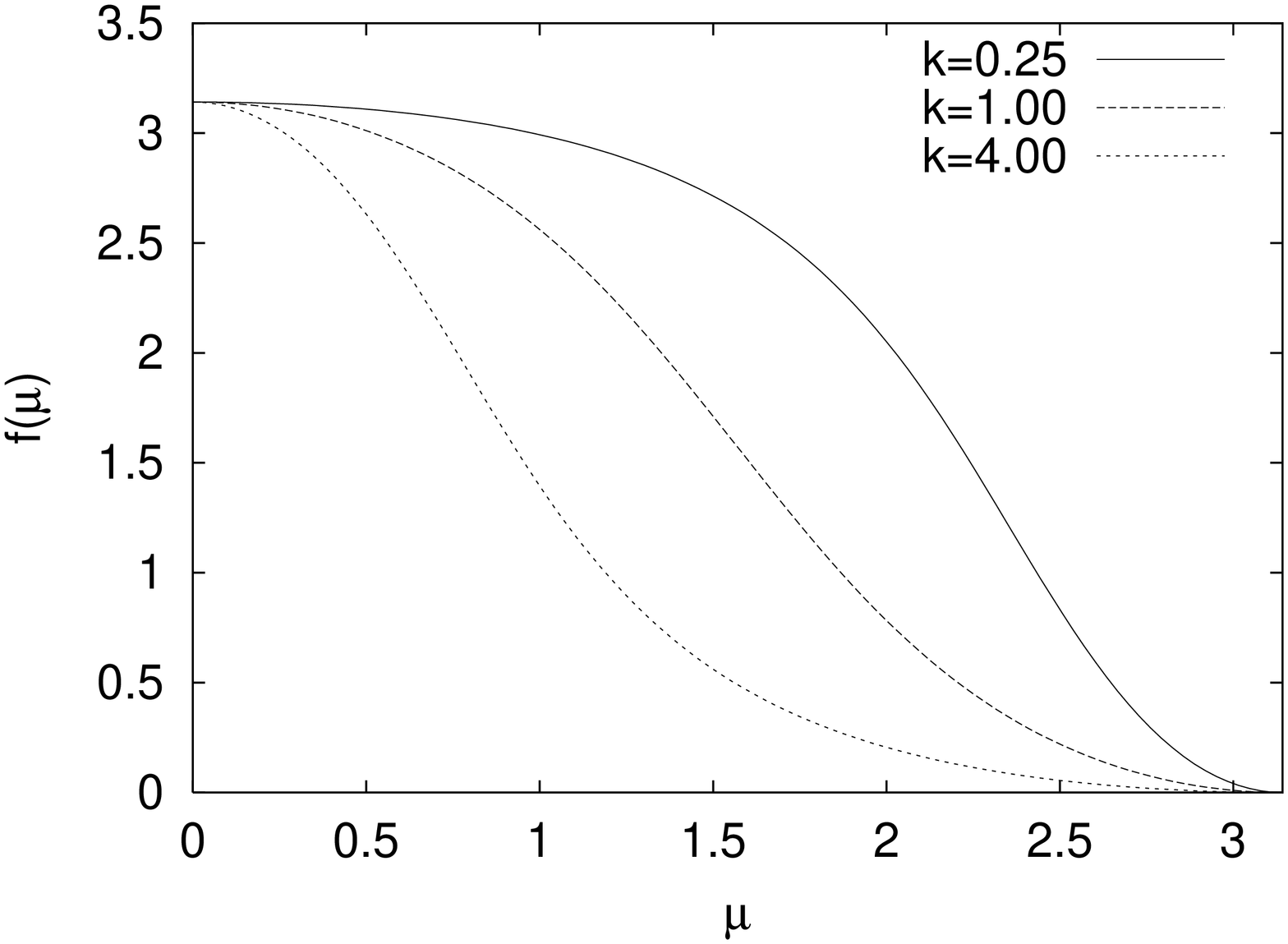}
}
\quad
\subfigure[$f^\prime(\mu)$ for $B=1$]{
\includegraphics[height=70mm,angle=270]{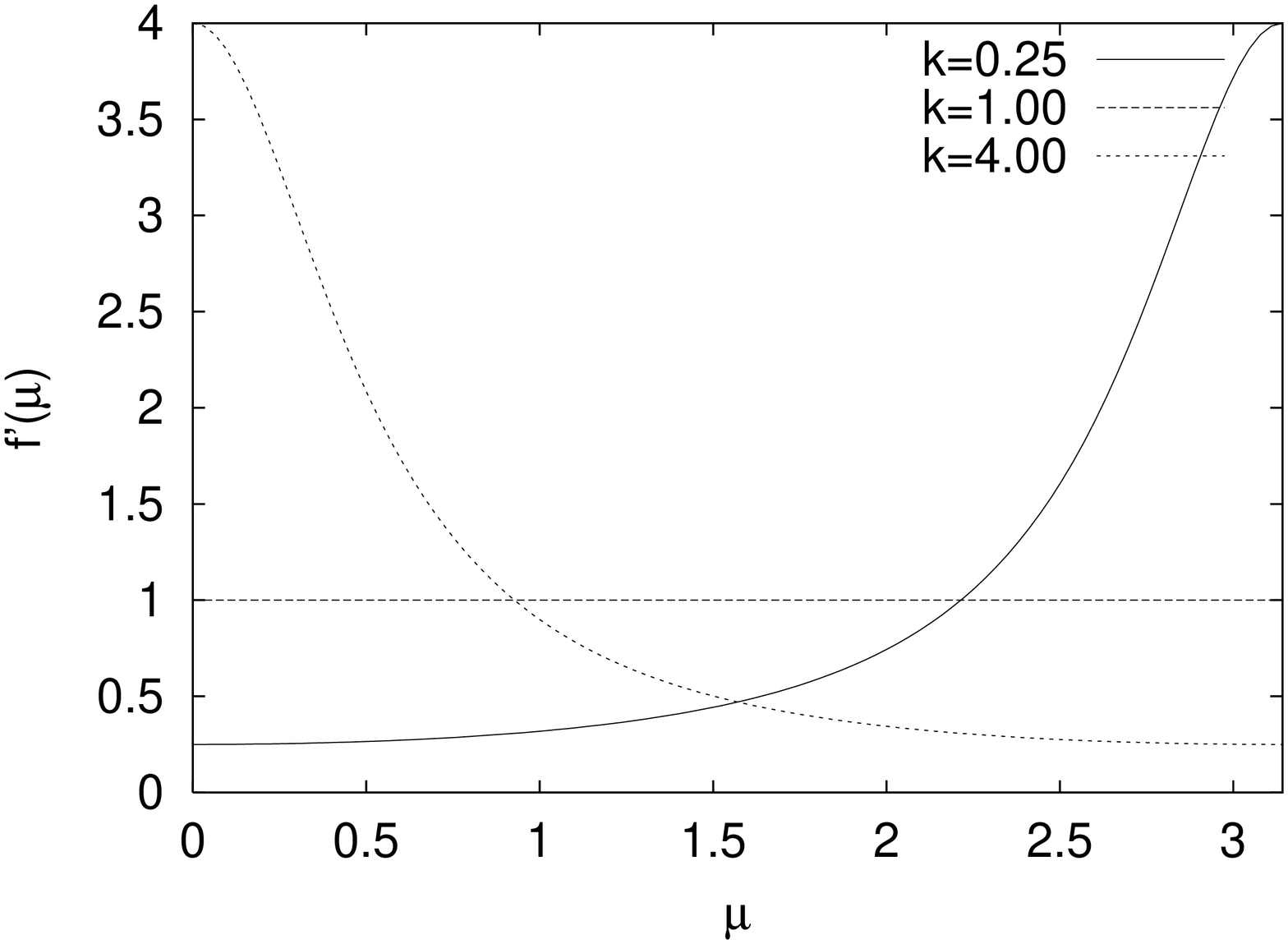}
}
\quad
\subfigure[$f^\prime(\mu)$ for $B=4$]{
\includegraphics[height=70mm,angle=270]{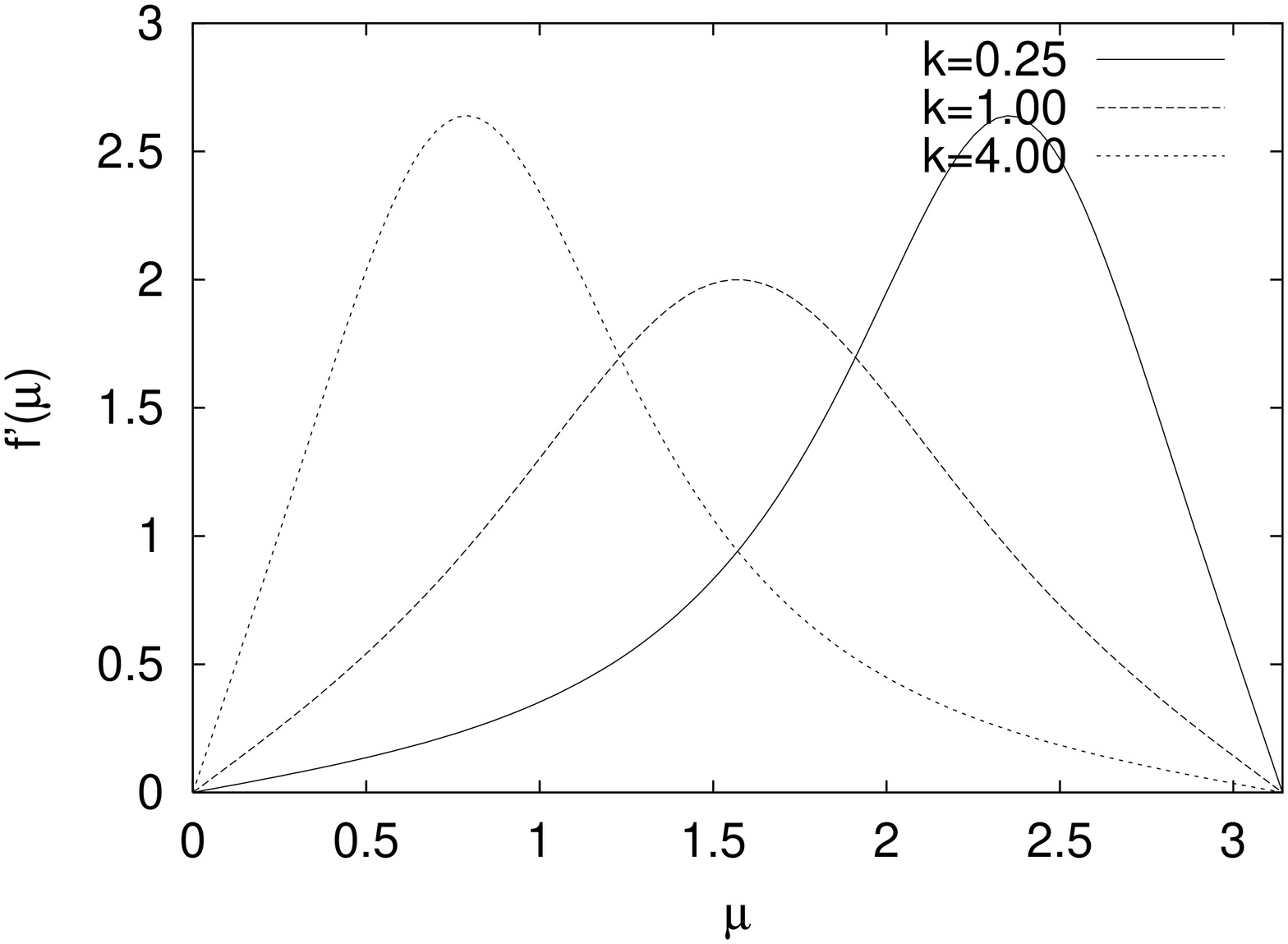}
}
\quad
\subfigure[${\tilde E}(\mu)$ for $B=1$]{
\includegraphics[height=70mm,angle=270]{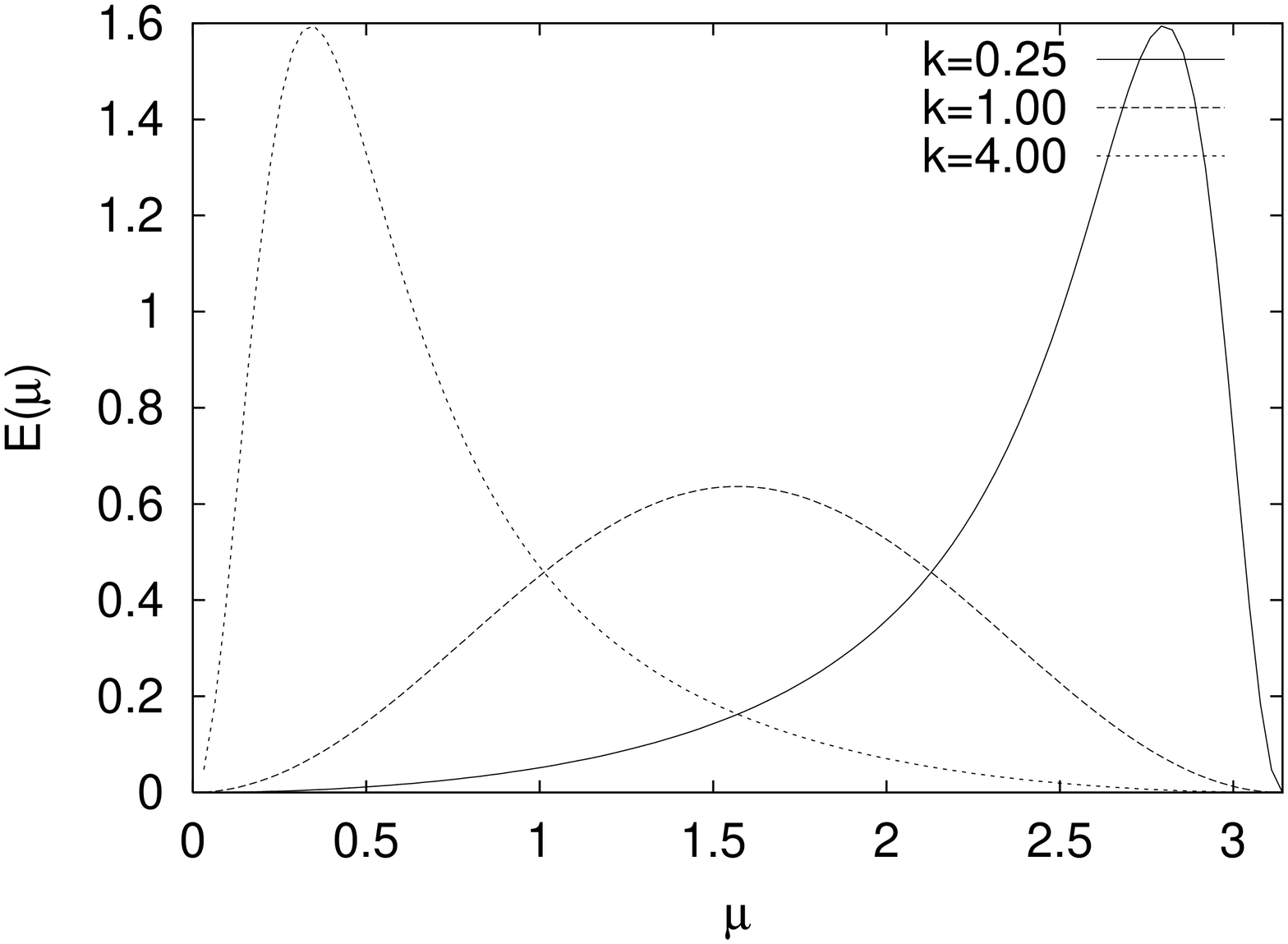}
}
\quad
\subfigure[${\tilde E}(\mu)$ for $B=4$]{
\includegraphics[height=70mm,angle=270]{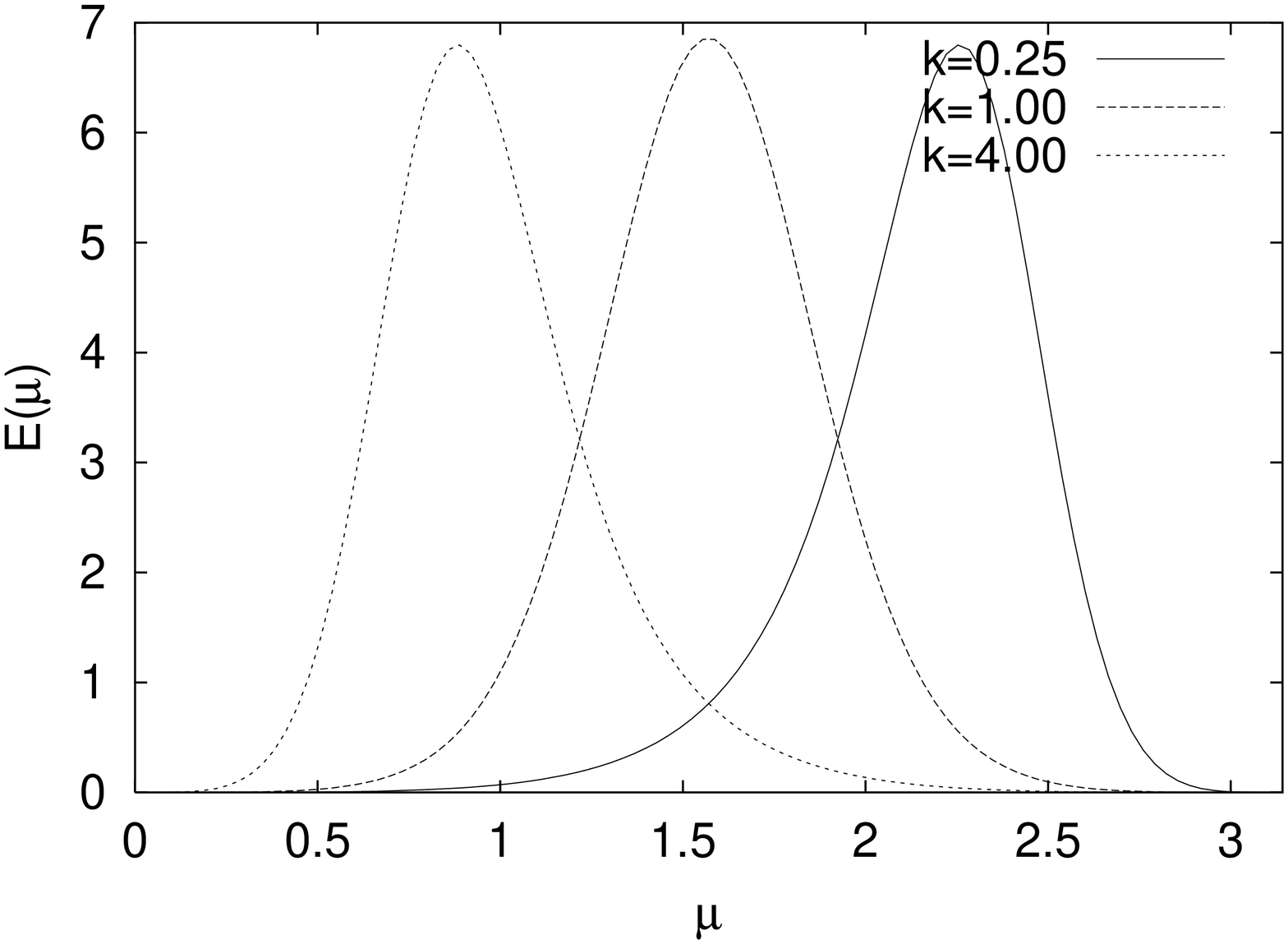}
}
\caption{The quasi-conformal shape function $f(\mu)$, its
derivative  $f^\prime(\mu)$, and the averaged energy density ${\tilde
E}(\mu)$ for $B = 1$ and $B = 4$.
The parameter $k$ takes the values $\frac{1}{4},1,4$. 
\label{fig1}} 
\end{center}
\end{figure}

Figure \ref{fig1} shows the shape function (\ref{confansatz}) and its
derivative for $B = N_R = 1, 4$ and for $k=\frac{1}{4}, 1, 4$. 
Figure \ref{fig1} also shows energy density ${\tilde E}(\mu)$ averaged
over the $2$-sphere, which is the integrand of (\ref{Erational}).
The shape function (\ref{confansatz}) 
has symmetry (\ref{fsym}) for $k=1$. For $k>1$ the
shape function is localized around the north pole $\mu = 0$,
whereas for $k<1$ it is localized around the south pole $\mu=\pi$. 
This is particularly obvious for its derivative $f^\prime(\mu)$.
Figure \ref{fig1} also
illustrates that for higher baryon number $B$ the symmetric solution
becomes more and more localized at the equator $\mu = \frac{\pi}{2}$.
Furthermore, the energy ${\tilde E}(\mu)$ has been plotted in order 
to compare the result to the usual Skyrme model. 
${\tilde E}(\mu)$ corresponds to an average radial density. 
When the energy is peaked close to $\mu=0$ this is the usual Skyrmion.
If the energy is centered around $\mu = \frac{\pi}{2}$ this
corresponds to a shell-like configuration.
Solutions centered around $\mu = \pi$ will correspond in the limit $L
\to \infty$ to a configuration centered around infinity, with an
infinite energy. In order to calculate ${\tilde E}(\mu)$ in figure \ref{fig1} 
we have to know the values for ${\cal I}$ and $L$.
As we shall see in the following, for a given $B$, 
$k \neq 1$ minimizes the energy (\ref{Eansatz}) for a unique radius $L$. 
$k=1$ is a solution for many different $L$, 
but among these solutions $L_{crit.}$ takes a special place.
For $B=1$, ${\tilde E}(\mu)$ has been calculated by setting
${\cal I} = 1$, and $L = 2.21$ corresponding to $k = 4$ and
$k=\frac{1}{4}$. For $k=1$ we have displayed ${\tilde E}(\mu)$ for the critical
length $L_{crit.}=1.41$.  
Similarly, for $B = 4$, $L = 2.47$ corresponds to $k=4$ and $k=\frac{1}{4}$, and
we have also displayed ${\tilde E}(\mu)$ for $L_{crit.} = 2.09$. 
${\cal I}$ takes the value ${\cal I} = 20.65$ 
of table \ref{tableA1} in the appendix. 

We will now discuss the energy (\ref{Eansatz}) as a function of $L$.
For $B=1$, and hence $N_R = 1$, 
the rational map is the identity map $R(z) = z$. The
integral (\ref{calI}) can be evaluated explicitly. We obtain ${\cal I}
= 1$ so that the energy (\ref{Eansatz}) is
\begin{equation}
 E = 12 \pi^2  \left( \frac{2 L}{k + \frac{1}{k} + 2} 
+ \frac{1}{4 L} \left( k + \frac{1}{k} \right) 
\right).
\end{equation}
It is convenient to set $\alpha =  k + \frac{1}{k}$ and to calculate
the minimum of the energy with respect to $\alpha$. 
$E$ has a single minimum at $\alpha_0 = \sqrt{8} L - 2$. For $L <
\sqrt{2}$, $\alpha_0$ is not attainable for real $k$, 
so that the minimum energy occurs at $\alpha = 2$ and $k=1$. 
In fact, up to a minus
sign this is the identity map between the physical $S^3_L$ and the target
$S^3_1$. The energy $E$ can now be written as
\begin{equation}
\label{E1k1}
E  = 6 \pi^2 \left( L + \frac{1}{L} \right).
\end{equation}
For $L > \sqrt{2}$ the minimum occurs where
\begin{equation}
\label{quadratic}
   k + \frac{1}{k} = \sqrt{8} L - 2.
\end{equation}
Equation (\ref{quadratic}) has two solutions. They are related by the
symmetry (\ref{fsymmetry}) and correspond to solutions localized at
one of the poles. Their energy is
\begin{equation}
\label{E1k2}
E = 12 \pi^2 \left( \sqrt{2} - \frac{1}{2 L} \right)
\end{equation} 
which is lower than (\ref{E1k1}). 
In the limit $L \to \infty$ the
energy becomes $E = 12 \pi^2 \sqrt{2}$, and the shape function turns into
$f(r) = 2 \arctan (\sqrt{8} r)$ where $r = L \mu$. 

\begin{figure}[!htb]
\begin{center}
\includegraphics[height=120mm,angle=270]{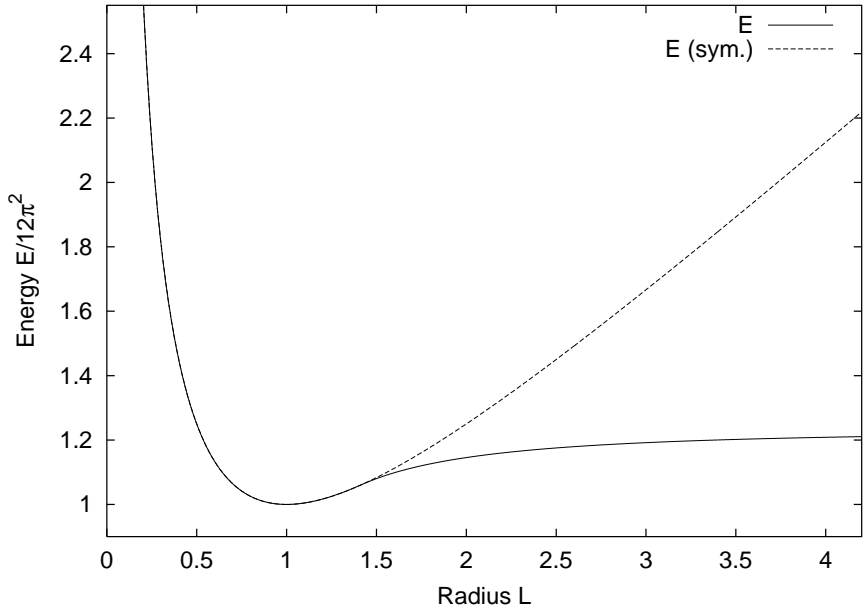}
\caption{The energy $E$ in units of $12 \pi^2$ of a $B=1$ Skyrmion as a
function of the radius $L$.\label{fig2}}
\end{center}
\end{figure}

Figure \ref{fig2} shows the energy $E$ of the Skyrmion as a function of
$L$. For small $L$ the energy scales like $\frac{1}{L}$. At the
optimal value of the energy $L_{opt.} = 1$ the energy reaches its
minimum $E_{opt.} =12 \pi^2$. 
The Skyrmion is symmetric if the radius $L$ is below the
critical radius $L_{crit.} = \sqrt{2}$.
For $L > L_{crit.}$ there are
two Skyrmions with the same energy. This phenomenon is often called
bifurcation.  
There is still a symmetric solution but it is no longer stable. 
For $L \to \infty$ the energy tends to the ${\mathbb R}^3$ value. 
For $B>1$ the energy $E$ shows similar behaviour.
The values $L_{opt.}$, $L_{crit.}$ and
$E_{opt.}$ depend on the baryon number $B$ and are good numbers to
characterize the behaviour of the solutions.

Now, we consider $N_R = 4$. 
In this case the optimal rational map has
octahedral symmetry and can be written in the following form:
\begin{equation}
R(z) = \frac{z^4 + 2\sqrt{3} i z^2 + 1}{z^4 - 2 \sqrt{3} i z^2 + 1}.
\end{equation}
We obtain ${\cal I} \approx 20.65$. Using expression
(\ref{Eansatz}) the energy is 
\begin{equation}
\label{EB4a}
E = \frac{24 \pi^2 \sqrt{2} L (\beta^3 - 6\beta + 4 \sqrt{2})}
{(\beta^2 - 2)^2}
+ \frac{5 \sqrt{2} \pi^2 (32 + {\cal I}) \beta}{16 L} 
\end{equation}
where $\beta = \sqrt{k} + \frac{1}{\sqrt{k}}$ plays the same role as
$\alpha$ for $B=1$. We can now minimize the energy with respect to $\beta$. This
gives rise to the cubic equation 
\begin{equation}
\label{eqcubic}
 (\beta -\sqrt{2})^3 - P (\beta -\sqrt{2}) - Q = 0,
\end{equation}
where $P = \frac{384 L^2}{5(32+{\cal I})}$ and 
$Q = \frac{768 \sqrt{2} L^2}{5 (32 + {\cal I})}$.
There is a real solution $\beta_0$ for all $L$ but its exact
form is long and not very illuminating.
For $L< L_{crit.}$ we have $\beta_0<2$ so that $\beta=2$ and $k = 1$
is again the minimum, and the solution is localized around the
equator.  
The critical length can be calculated by substituting $\beta = 2$ into
equation (\ref{eqcubic}), and we obtain
\begin{equation}
\label{B4Lcrit}
L_{crit.} = \frac{1}{24} \sqrt{\frac{15 (32+{\cal I})}{8 \sqrt{2} -11}}
~~~\approx 2.091
\end{equation}
For $k=1$ the energy simplifies:
\begin{equation}
\label{E4k2}
E =  24 \pi^2 L (2 -\sqrt{2})
+ \frac{5 \sqrt{2} \pi^2 (32 + {\cal I})}{8 L}.
\end{equation}
The optimal length can be calculated by minimizing $E$ with
respect to $L$:
\begin{equation}
L_{opt.}  = \sqrt{\frac{5 \sqrt{2}(32 + {\cal I})}
{192(2 -\sqrt{2})}} ~~~\approx 1.819.
\end{equation}
For $L>L_{crit.}$ there are two solution such that $\sqrt{k} +
\frac{1}{\sqrt{k}} = \beta_0$ related by the symmetry
(\ref{fsymmetry}). The corresponding expression for the energy can be
derived explicitly by setting $\beta = \beta_0$ in equation (\ref{EB4a}) 
but it is rather lengthy.

The limit $L \to \infty$ can be derived from equation
(\ref{eqcubic}). We are interested in the solution where $\beta$ is
of order $L$ such that the energy $E$ in equation (\ref{EB4a}) is
finite. Therefore, we can neglect the last term in (\ref{eqcubic}),
and the $\sqrt{2}$, and obtain $\beta = \sqrt{P}$. Inserting $\beta$
into the expression for the energy (\ref{EB4a}) we obtain
$\frac{E}{12 \pi^2} = 4.684$ which corresponds to $1.171$ per baryon.
		 
For $N_R = 9$ the optimal rational map has $D_{4d}$ symmetry and ${\cal I}
\approx 109.3$. 
It is again possible to introduce an auxiliary variable
$\gamma = k^\frac{1}{3} + k^{-\frac{1}{3}}$. However, minimizing the
energy leads to a polynomial of degree $5$ in $\gamma$. Therefore it is
no longer possible to calculate the solution explicitly. But the
general behaviour remains the same. The solution is symmetric below a
critical length $L_{crit.}$  which is determined by substituting $\gamma =
2$ into the equation for minimizing the energy. We obtain 
\begin{equation}
\label{B9Lcrit}
L_{crit.} = \frac{2}{513} \sqrt{323190 + 1195~ {\cal I}} 
~~~ \approx 2.887.
\end{equation}
The optimal radius can also be calculated analytically:  
\begin{equation}
L_{opt.} =  \frac{1}{297} \sqrt{374220 + 2310~ {\cal I}}~~~
\approx 2.683.
\end{equation}
For $L>L_{crit.}$ there are again two degenerate solutions.
In the limit $L \to \infty$ we can balance the terms which are of
highest order in $L$. This leads to $\gamma \approx 0.779 L$, and the 
normalized energy is $\frac{E}{12 \pi^2} = 10.274$ which is $1.142$
per baryon.
         
In the last part of this section we will discuss the limit $L \to
\infty$ for general $N_R$. 
Setting $r = L \mu$ the quasi-conformal ansatz
(\ref{confansatz}) tends to
\begin{equation}
\label{flatansatz}
f(r) = \pi - 2 \arctan \left( \left( \frac{r}{R_0} 
\right)^{\sqrt{N_R}} \right)
\end{equation}
where $R_0$ is a free parameter. The energy can be written as 
\begin{equation}
\label{ER0}
E = \frac{4}{3 \pi} \left(
\frac{3 N_R}{R_0} I_1 + \left(2 N_R^2 + {\cal I} \right) R_0 I_2 
\right)
\end{equation}
However, now the integrals $I_1$ and $I_2$ are independent of $R_0$
and depend only on $N_R$:
\begin{eqnarray}
I_1 &=& \int_0^\infty \frac{r^{2\sqrt{N_R}}}{(1+r^{2 \sqrt{N_R}})^2} 
{\rm d}r \\
\nonumber \\
I_2 &=& \int_0^\infty \frac{r^{4\sqrt{N_R}}}{r^2 (1+r^{2
\sqrt{N_R}})^4} 
{\rm d}r
\end{eqnarray}
It is straightforward to minimize the energy (\ref{ER0}) 
with respect to $R_0$. In figure \ref{fig6} and \ref{fig7}, 
in the following section
we have evaluated the minimal energy $E(R_{0} = R_{0,min})$, and the
optimal radius $R_{0,min}$,
respectively.
The parameter $R_{0,min}$
has a natural interpretation as the size of the Skyrmion: 
For $r = R_{0,min}$ the shape function $f(r)$ has reached the value
$\frac{\pi}{2}$. Furthermore, when the baryon density is integrated
over a ball of radius $R_{0,min}$, then this is just half the total baryon
number:
\begin{equation}
\frac{2 N_{R}}{\pi} \int_{0}^{\frac{\pi}{2}}
\sin^2 f~ {\rm d} f = \frac{B}{2}.
\end{equation}
It is also worth noting that equations (\ref{confansatz}) for $k=1$
and (\ref{flatansatz}) are related by $r = R_{0,min} \tan \frac{\mu}{2}$. 
This is a stereographic projection from a $3$-sphere of radius $R_{0,min}$
to ${\mathbb R}^3$.  
Minimizing the energy we see that a Skyrmion in flat
space is related to a Skyrmion on the $3$-sphere with an optimal
radius $L_{opt.}$. 

\section{Numerical Results}
\label{NumericalResults}

In this section we discuss the numerical solution of equation
(\ref{Euler}) for the shape
function $f(\mu)$ of the rational map ansatz and the properties of the
resulting fields, and we compare it to the analytical solutions from
the previous section.
Here, we only
consider configurations with $N_f=1$ such that $B=N_R$. A
configuration with $N_f = 2$ is discussed in the following section.

\begin{figure}[!htb]
\begin{center}
\includegraphics[height=120mm,angle=270]{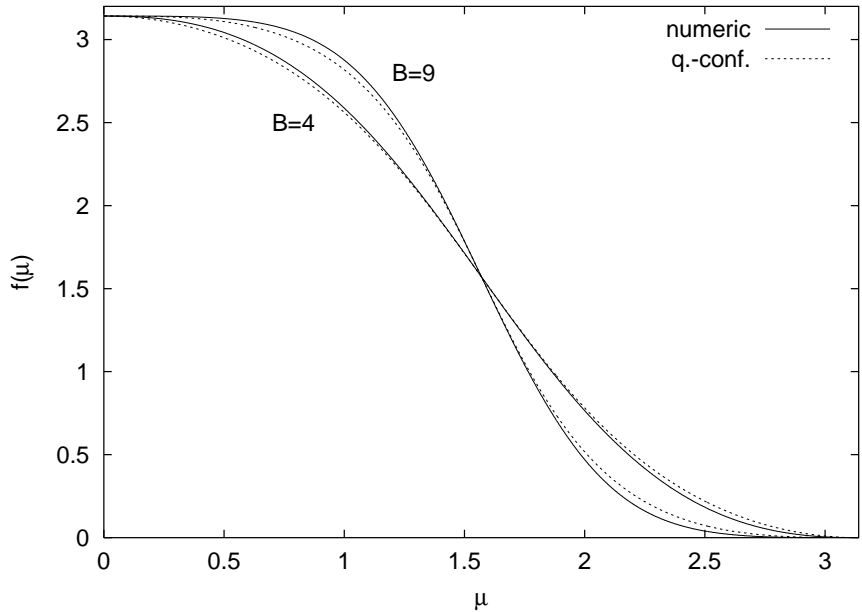}
\caption{Comparison of the numerical shape function with the
quasi-conformal ansatz for $B=4$ and $B=9$. \label{fig3}}
\end{center}
\end{figure}

The numerical solution of (\ref{Euler}) is calculated using a
relaxation method.  
For symmetric initial conditions we obtain the symmetric solution
which is a saddle point for $L > L_{crit.}$. 
For asymmetric initial conditions we obtain the minimum energy configuration,
{\it i.e.} the Skyrmion.  
In figure \ref{fig3} we compare the numerical shape function
with the quasi-conformal ansatz (\ref{confansatz}) for $B=4$ and $B=9$
at the optimal radius $L=L_{opt.}$. The optimal radius $L_{opt.}$
implies $k=1$ in the quasi-conformal ansatz. The numerical result and the
quasi-conformal ansatz show good agreement, in particular in the
region around the equator $\mu = \frac{\pi}{2}$. This region is most
important for the energy (\ref{Erational}) since both $(f^\prime (\mu))^2$
and $\sin^2 f(\mu)$ are large for $\mu \approx \frac{\pi}{2}$. Near the
poles the numerical solution is less steep than the quasi-conformal
ansatz as we expect from their exponents:  $\rho_+ > \sqrt{B}$.

\begin{figure}[!htb]
\begin{center}
\includegraphics[height=160mm,angle=270]{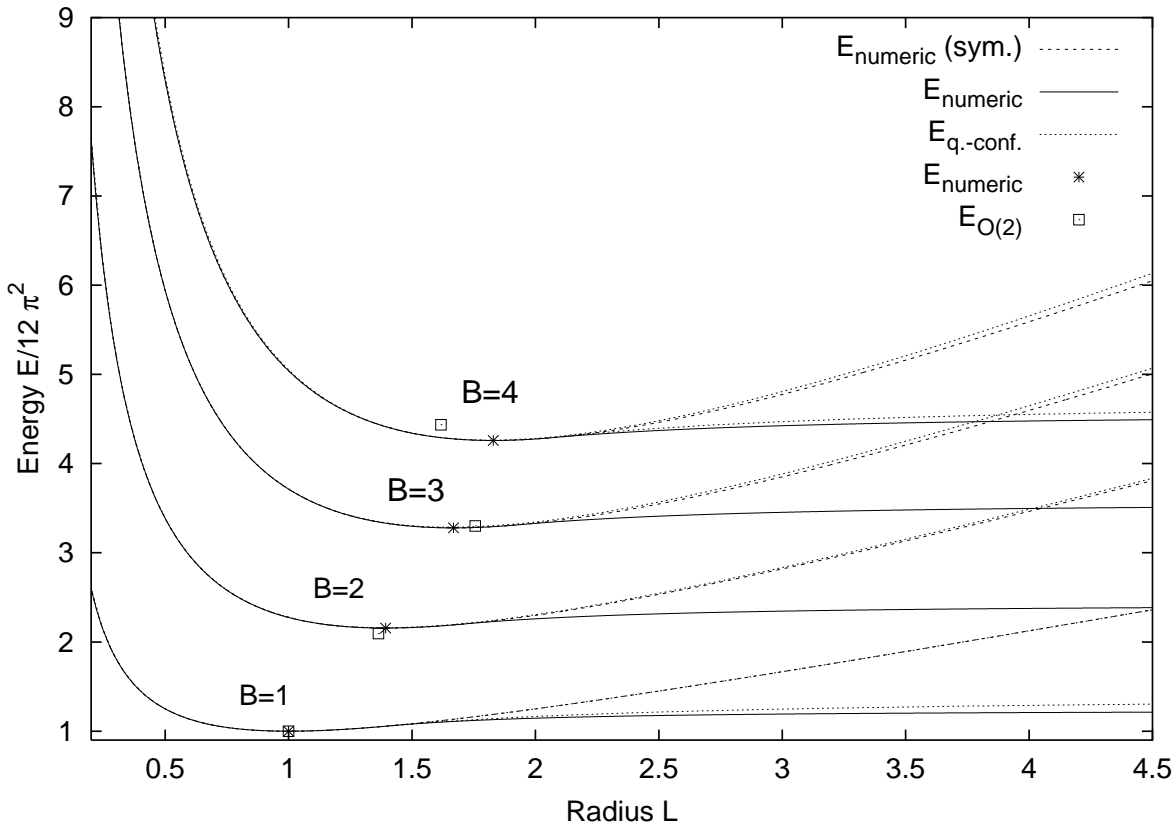}
\caption{Comparison of the numerical calculation and the
quasi-conformal ansatz for the energy $E$ as a function of the radius
$L$ for baryon number $B=1,\dots,4$.\label{fig4}} 
\end{center}
\end{figure}

\begin{figure}[!htb]
\begin{center}
\includegraphics[height=120mm,angle=270]{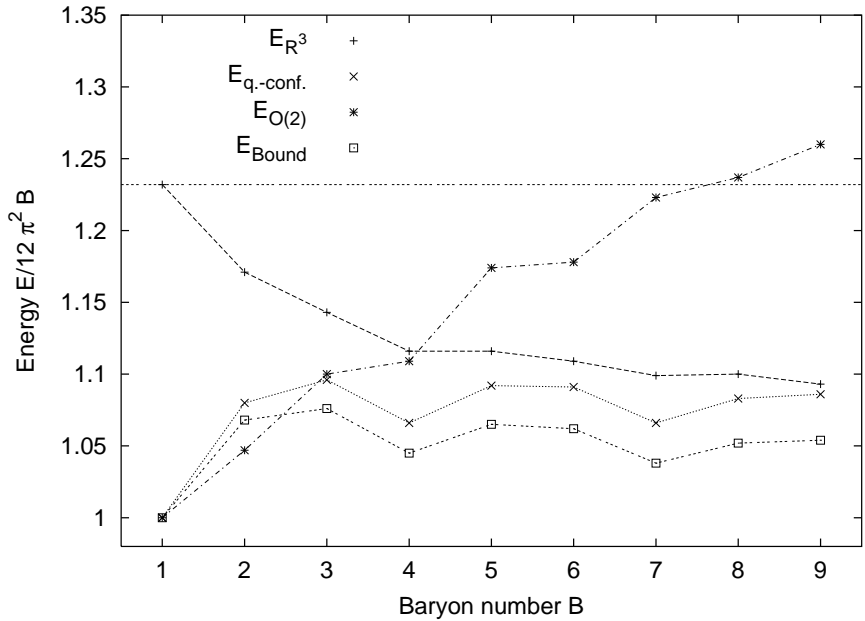}
\caption{The energies $E_{opt.}^{q.-conf.}$, $E_{opt.}^{O(2)}$, 
$E_{{\mathbb R}^3}$ and
$E_{Bound}$ per baryon 
as a function of the baryon number $B$.\label{fig5}}
\end{center}
\end{figure}

In figure \ref{fig4} we show the energy $E$ as a
function of $L$ for $B= 1,\dots,4$ and compare it to our analytic
calculations of the previous section. 
In all the figures the energy $E$ is normalized by $12 \pi^2$ so that the
Faddeev-Bogomolny bound (\ref{Bogbound}) is equal to $|B|$.
For $B=1$ we plot energy (\ref{E1k1}) of the symmetric solution 
which gives the exact result. 
For $L> L_{crit.}=\sqrt{2}$ we also plot the energy (\ref{E1k2}). Our
ansatz predicts the correct critical radius but energy (\ref{E1k2}) is
slightly too high.
For $B = 2$ and $B=3$ we only display the energies of the
symmetric solutions using formula (\ref{Eansatz}) and setting
$k=1$.
The integrals $I_1(1)$ and $I_2(1)$ in (\ref{Eansatz}) are calculated
numerically. Again there is very good agreement for small $L$, 
and for $L>L_{crit.}$ the ansatz is also very close to the
symmetric solution.
For $B = 4$ we plot the analytic expression for the symmetric solution
(\ref{E4k2}).  
For $L>L_{crit.}^{q.-conf.}\approx 2.09$ we also plot
(\ref{EB4a}) with $\beta = \beta_0$. The critical radius
$L_{crit.}^{q.-conf.}$ is slightly higher than $L_{crit.}^{numeric} =  2.071$, and
the energy is also slightly too high. 
In figure \ref{fig4} we also mark the minimal energy
$E_{opt.}^{numeric}$ for the optimal radius $L_{opt.}^{numeric}$ and for
comparison the optimal energy $E_{opt.}^{O(2)}$ for the doubly 
axially-symmetric ansatz at its optimal radius $L_{opt.}^{O(2)}$. We will
discuss these quantities in figure \ref{fig5}.

In summary, all the analytical results are in good agreement with the
numerical solution for a large range of $L$. 
This is quite remarkable given that the ansatz for the
shape function only depends on one parameter $k$. 

In figure \ref{fig5} we compare the optimal energy of the
quasi-conformal map $E_{opt.}^{q.-conf.}$ with the energy of the doubly 
axially-symmetric ansatz $E_{opt.}^{O(2)}$ for all baryon numbers $B=1,\dots,9$.
For $B=1$ both reproduce the exact result. For $B=2$ the doubly
axially-symmetric ansatz is believed to be the exact solution because
its symmetry is compatible with the results in flat space. 
Yet, for $B>2$ the energy $E_{opt.}^{q.-conf.}$ is lower than $E_{opt.}^{O(2)}$. 

Figure \ref{fig5} also shows that the quasi-conformal ansatz always
has a lower energy than the numerical solution in ${\mathbb R}^3$. We
conclude that $L_{opt.}$ for the exact solution will be finite.
The horizontal line at $E=1.232$ in figure \ref{fig5} corresponds to $B$ well
separated single Skyrmions. For $B \ge 8$ the energy of the doubly
axially-symmetric ansatz is above this line which is unphysical. 
Note that the energy per baryon in flat space $E_{{\mathbb R}^3}$
decreases as we increase the baryon number. 
We would obtain a similar behaviour if we kept the radius $L$
fixed. Yet, we then have to make the radius $L$ large enough so
that the largest Skyrmion ``fits''. 
The physical interpretation is that one multi-Skyrmion is more stable
than a number of smaller Skyrmions. 

We also display the rational map bound $E_{Bound}$ (\ref{Bograt}). 
For $B=2$ the energy of the doubly axially-symmetric ansatz is
lower than $E_{Bound}$. This confirms our warning that energies of 
exact solutions do not have to be greater than $E_{Bound}$. 
However, figure \ref{fig5}
shows that the bound follows the general behaviour of
$E_{opt.}^{q.-conf.}$ and also of $E_{{\mathbb R}^3}$.

\begin{figure}[!htb]
\begin{center}
\includegraphics[height=120mm,angle=270]{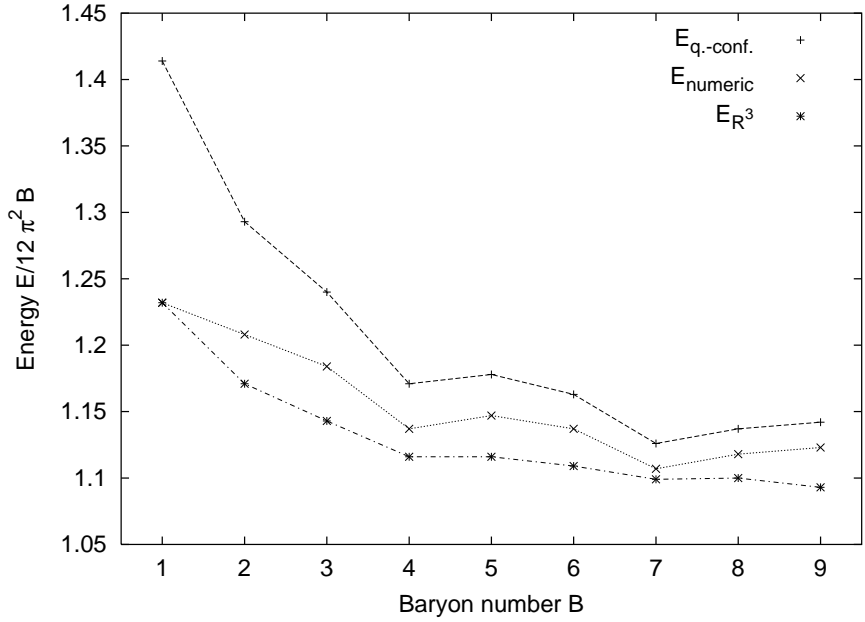}
\caption{The energy $E_{{\mathbb R}^3}$ and the energies 
$E_{opt.}^{q.-conf.}$ and 
$E_{opt.}^{numeric}$ per baryon for $L \to \infty$ as a function of
$B$.\label{fig6}}  
\end{center}
\end{figure}

In figure \ref{fig6} we compare the energy  of the
quasi-conformal map $E_{opt.}^{q.-conf.}$  in the limit $L \to \infty$ 
to the rational map energies
$E_{opt.}^{numeric}$ and the energy $E_{{\mathbb R}^3}$ of the exact solution 
in
flat space.
$E_{opt.}^{q.-conf.}$ is calculated by minimizing the energy of expression
(\ref{ER0}) with respect to $R_0$. For $B = 1,4,9$ the energy
$E_{opt.}^{q.-conf.}$ agrees with the results of section \ref{ConformalMap}.
Figure \ref{fig6} shows that the quasi-conformal ansatz approximates the
numerical solution well. The relative difference between $E_{opt.}^{q.-conf.}$
and $E_{opt.}^{numeric}$ is monotonically decreasing and is less than $3\%$ for
$B=4$. This means that the error of the rational map ansatz with a
numerical shape function and the quasi-conformal map ansatz are of the
same order of magnitude. 

\begin{figure}[!htb]
\begin{center}
\includegraphics[height=120mm,angle=270]{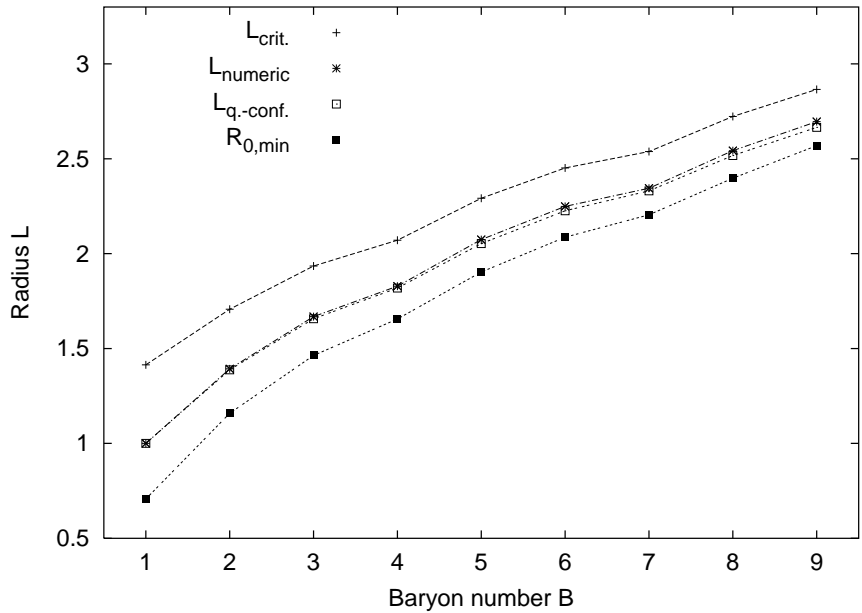}
\caption{The radii $L_{opt.}^{numeric}$, $L_{opt.}^{q.-conf.}$, $L_{crit.}$,
and $R_{0,min}$ as a function of the baryon number $B$.\label{fig7}}
\end{center}
\end{figure}

In figure \ref{fig7} we compare the different optimal radii. 
The optimal radius of the quasi-conformal ansatz $L_{opt.}^{q.-conf.}$ 
is always slightly smaller than $L_{opt.}^{numeric}$.  
In all cases the value of the critical radius $L_{crit.}$ is greater
than the optimal radius so that the results are consistent. 
The critical radius can be calculated in
two different ways. In the first approach we determine the point at
which the solution bifurcates. 
We use the fact that the symmetric initial condition
relaxes into the symmetric solution which is always present, 
whereas an asymmetric initial condition only relaxes to the symmetric
solution when there is no asymmetric solution. This method leads to
rather large error bars because it is difficult to decide whether two
numerically calculated solutions are different.
The second approach is to consider the value of the following
integral: 
\begin{equation}
\label{intf}
\int_0^\pi \left( f(\mu) + f(\pi-\mu) - \pi \right)^2 {\rm d} \mu.
\end{equation}  
This integral is zero if the solution possesses the symmetry
(\ref{fsymmetry}) and is nonzero otherwise. 
We determine the value of $L$ where the integral
(\ref{intf}) first becomes nonzero. Both approaches give the same
results, but the second one is much more accurate. At the critical
value there is a jump in the value of the integral (\ref{intf}). 
This is also a verification of our analytic result that the solution
is symmetric for $L < L_{crit.}$.
As mentioned before, for $B=1$ the
critical radius is in agreement with the analytic result $L_{crit.} =
\sqrt{2}$. For $B = 4$ the critical value of the quasi-conformal ansatz
$L_{crit.}^{q.-conf.} = 2.091$ in (\ref{B4Lcrit}) is
only slightly higher than the critical radius of the rational
map $L_{crit.}^{numeric} = 2.071$. Similarly, for $B=9$ we calculated
$L_{crit.}^{q.-conf.}  = 2.887$ in (\ref{B9Lcrit}) whereas
$L_{crit.}^{numeric} = 2.866$.  

In figure \ref{fig7} we also find that $R_{0,min}$ is always smaller
than $L_{opt.}^{q.-conf.}$. The difference between these lengths becomes
smaller the larger the baryon number is. This confirms our
interpretation stated at the end of the previous section, that the Skyrmion
in flat space is related to the Skyrmion on a $3$-sphere with optimal
radius.

\section{Phase Transitions}
\label{Phasetransitions}

In this section we will describe phase transitions on
$S^3$ as $L$ is varied. 
There is no numerical work solving the full set
of coupled partial differential equations for $S^3_L$ as there is in
flat space.
Therefore, we have to rely on our ans\"atze. First, we
will discuss two different order parameters. Then we will describe 
particular situations in more detail.

In ${\mathbb R}^3$ the average $\langle \sigma \rangle$ is
equal to $1$ because of the boundary conditions at infinity. 
However, on $S^3_L$ it is easy to show that
$ \langle \sigma \rangle$  vanishes for $L < L_{crit.}$ when the
solution is symmetric with respect to south and north pole. 
Therefore, we can
choose $O_1 = \langle \sigma \rangle^2$ as an order
parameter.\footnote{A more symmetric choice is $\langle \sigma
\rangle^2 + \langle \pi_i \rangle^2$, \cite{Jackson-89}.}
It has the necessary property that it is nonzero above the phase
transition $L > L_{crit.}$ and identically zero below the transition.  
As we already pointed out in section \ref{RationalMaps}
this order parameter also vanishes identically for all doubly 
axially-symmetric configurations. This is consistent with the fact that for
$L>L_{crit.}$ these configurations are bad approximations.  
Another possible order parameter would be $ O_2 = \langle \sigma^2 \rangle
-\frac{1}{4}$. This parameter $O_2$ is motivated by the fact that enhanced
chiral symmetry, that is a symmetry that mixes $\sigma$ and $\pi_i$,
could result in a vanishing of $O_2$.

For $B=1$ the situation is well understood. Both parameters $O_1$
and $O_2$ vanish at the same time. Physically, this means that the
localization--delocalization transition occurs at the same time as
the restoration of chiral symmetry \cite{Manton-87}. 
In fact, for $L < \sqrt{2}$ the solution possesses full chiral
symmetry $SO(4)$.  For $B>1$ the chiral symmetry will be a subgroup of
$SO(4)$.

\begin{figure}[!htb]
\begin{center}
\subfigure[The energy of $B=4$ Skyrmions with the rational
map ansatz and with the doubly axially-symmetric
ansatz.\label{fig8a}]{ 
\includegraphics[height=75mm,angle=270]{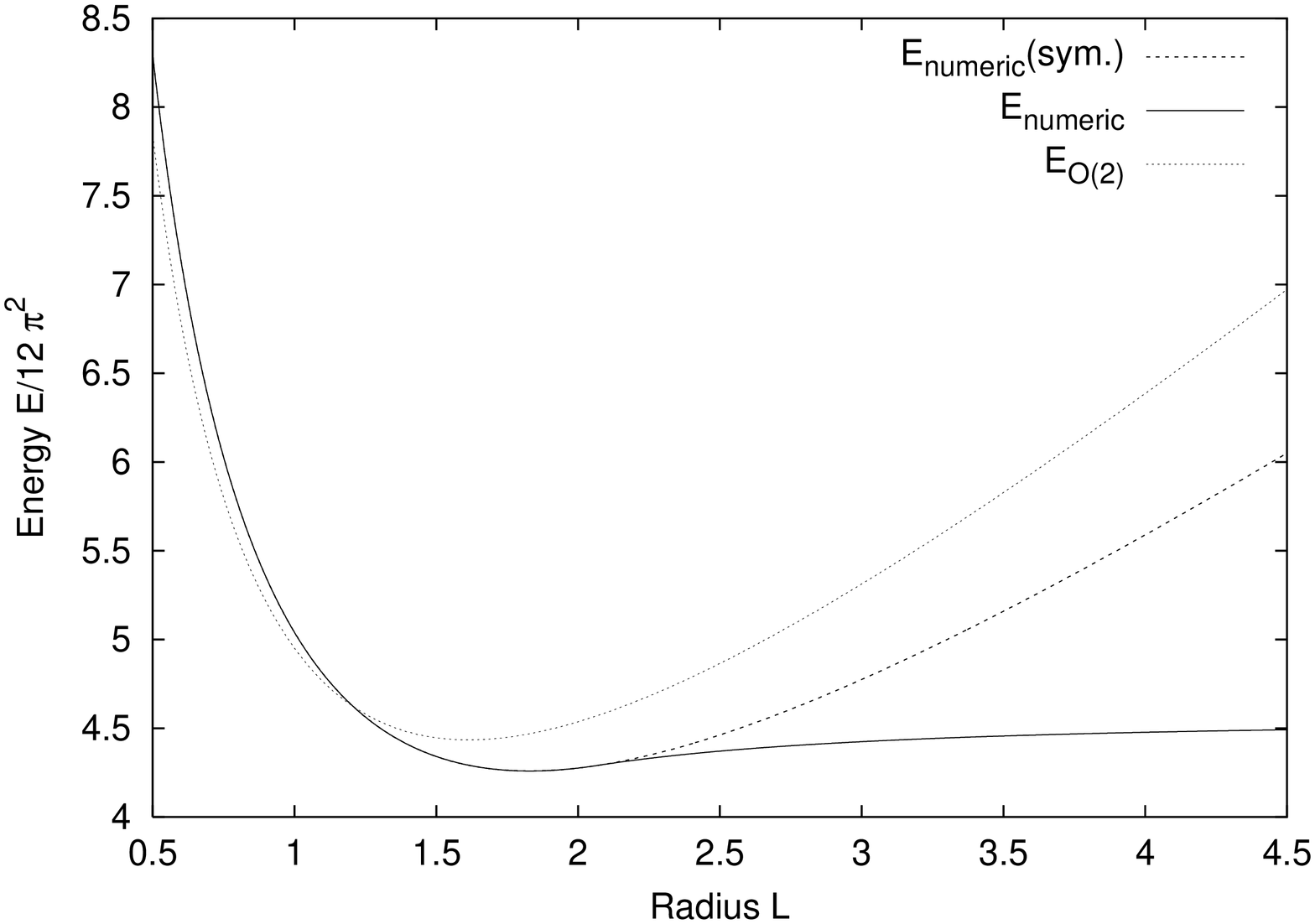}
}
\quad
\subfigure[The energy of two $B=8$ Skyrmions, one with $N_f=1$ and one
with $N_f = 2$.\label{fig8b}]{
\includegraphics[height=75mm,angle=270]{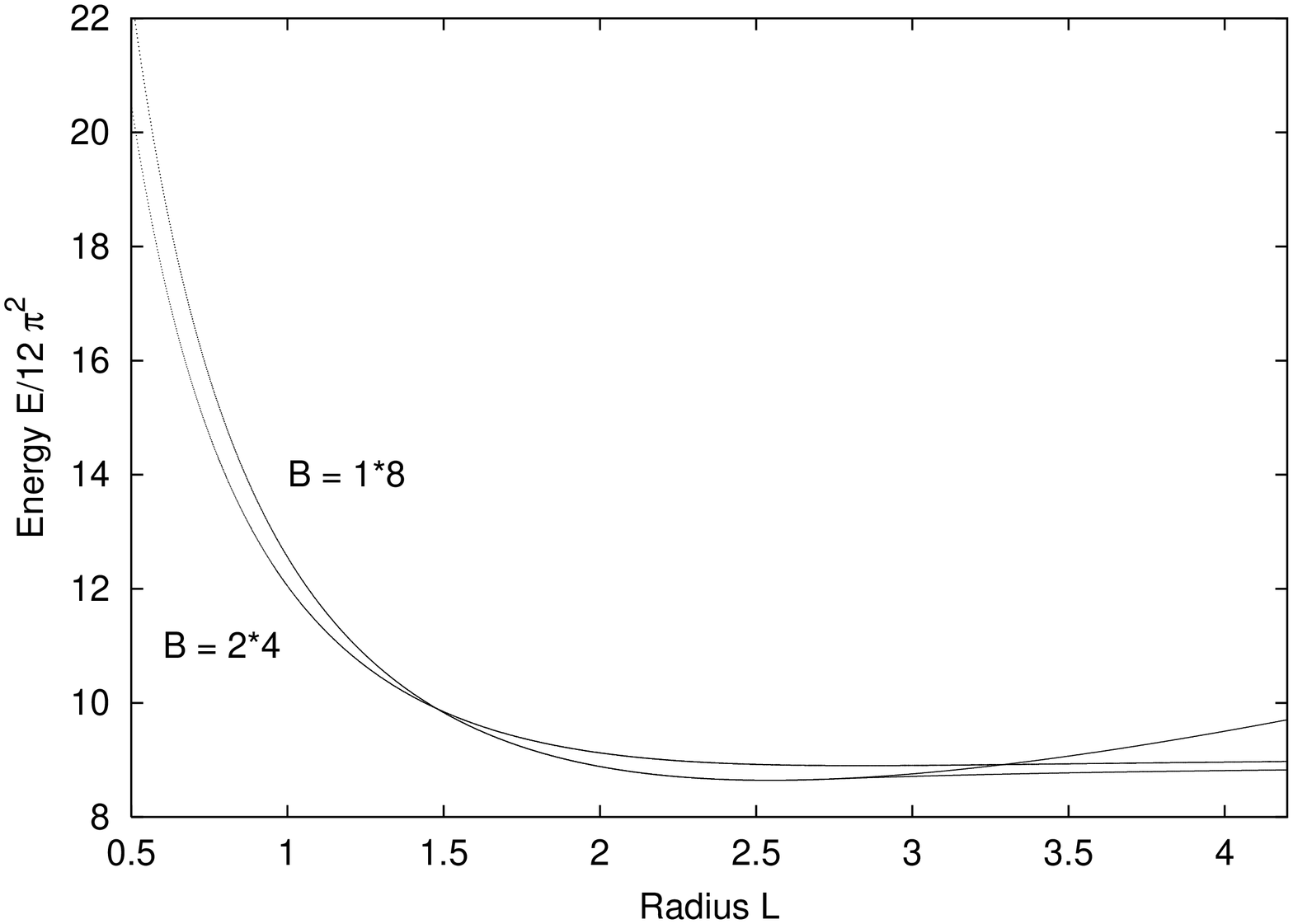}
}
\caption{Phase transitions in the Skyrme model}
\end{center} 
\end{figure}

Next we discuss $B=4$ and use both the rational map ansatz and the
doubly axially-symmetric ansatz. In figure \ref{fig8a} we plot their
energies. For $L > L_{crit.}$ the Skyrmion is localized at one of the
poles. For smaller $L$ the Skyrmion is localized around the
equator. This is the first phase transition and the order parameter is
$O_1$. 
 However, there is a radius $L_{crit.^\prime}$ where the doubly
axially-symmetric ansatz with $p=2$ and $q=2$ becomes the minimal
energy solution. As pointed out in section \ref{RationalMaps} if $p=q$
the solutions have the symmetry (\ref{gsymmetry}). For this
special case we checked numerically that the symmetry is actually a
symmetry of the shape function $g(\chi)$:
\begin{equation}
g\left(\chi \right) = \frac{\pi}{2} - g\left(\frac{\pi}{2} - \chi \right)
\end{equation}
This gives rise to a symmetry between $\sigma$ and  $\pi_1$ and
leads to the vanishing of the order parameter $O_2$. Therefore, we
have a second phase transition at $L = L_{crit.^\prime}$.

As a last example we describe a Skyrmion with baryon number $B=8$. 
For large $L$ the $N_f=1$ solution
has minimal energy. However, there is a saddle point solution where two
$B=4$ Skyrmions are located at opposite poles of the $3$-sphere.
If $L$ is sufficiently small there is a solution with $N_f=2$
which has lower energy than the one with $N_f=1$ as shown in figure
\ref{fig8b}.\footnote{Table \ref{tableA2} in the appendix 
gives an indication when to
expect a configuration with $N_f>1$ to be more stable than the
corresponding configuration with $N_f = 1$.} 
This configuration is particularly interesting because both
$B=4$ Skyrmions have cubic symmetry. For sufficiently small radius
$L_{hc}$ these two Skyrmions could combine and form a solution with 
a hypercubic symmetry. 
This symmetry would be strong enough to restore the symmetry
between the $\sigma$- and $\pi_i$-fields. The order parameter
$O_2$ would vanish.
In this scenario there
would be a localization--delocalization transition at $L_{crit.}$,
whereas the chiral symmetry is further enhanced at
$L_{hc}$. Unfortunately, these discrete subgroups of the full $SO(4)$
symmetry cannot be studied with the rational map ansatz.

In physical terms we propose the following picture.
Reducing the radius $L$
increases the baryon density. If we start with a $B >1$ Skyrmion at
large length this is well described by the rational map ansatz. There
is a phase transition at the critical radius $L_{crit.}$ where the
Skyrmion becomes delocalized over the $3$-sphere and the order
parameter $O_1$ vanishes. Then, one or more
phase transitions follow which make the configuration more and more
symmetric, and chiral symmetry is partially restored. Note however, the
Skyrme model is an effective theory which is only valid for low
energies. If the radius is too small we expect that further terms in
the effective action become physically important.

\section{Conclusion}
\label{Conclusion}

In this paper we have described the rational map ansatz applied to
Skyrmions on a $3$-sphere of radius $L$. 
We have calculated the energy $E$ of a Skyrmion as a function of $L$
for baryon number $B=1,\dots,9$ and found the following  behaviour: 
For small $L$ the energy $E$ scales like $\frac{1}{L}$. 
$E$ has a global minimum at an optimal radius
$L_{opt.}$. When
the radius $L$ is further increased there is a bifurcation point at a
critical length $L_{crit.}$:
below $L_{crit.}$ the Skyrmion is symmetric
under reflection at the plane through the equator of $S^3$. 
Above $L_{crit.}$ there are two degenerate Skyrmions one of 
which is localized at the north pole whereas the other one is
localized at the south pole. 
For $L \to \infty$ the ansatz
tends to the rational map ansatz for ${\mathbb R}^3$, \cite{Houghton-98}. 
For $B=1$ and $L=1$ our ansatz reproduces the known exact solution
\cite{Manton-Ruback}.
For $B=2$ our results are worse for small $L$ than the doubly
axially-symmetric ansatz in \cite{Jackson-89}. 
However, for $B > 2$ the energies of our solutions are lower than any
solutions known to date, including the doubly axially-symmetric
solutions. 

We have also derived an analytic expression for the shape function by
imposing that the metrics on the physical $S^3$ and on the target
$S^3$ are conformal in an average sense. 
This quasi-conformal ansatz for the shape function depends only on one
parameter $k$.  By varying $k$ we obtain very good  
agreement with the numerical results for a large range of $L$,
even for flat space.   
It is worth noting that the relative error of the quasi-conformal ansatz decreases
with increasing baryon number. 
We have shown that the solution is localized at the equator if $k=1$.
For $B=1,4,9$ we have shown that the Skyrmion really is
symmetric below a critical value $L_{crit}$, {\it i.e.}  $k=1$ gives
the minimal energy solution.  

One particular property of the quasi-conformal shape function is that it
becomes more and more localized around the equator as the baryon number
increases. To be more precise, the derivative of the shape
function, which is connected to the baryon density, has a peak at the equator. 
Since this ansatz was derived with the assumption that the
map between the metrics is ``as conformal as possible'' we have found a
geometric explanation of why Skyrmions on the $3$-sphere are
shell-like. 
In flat space the map cannot be conformal.  However, one ``half'' of
the Skyrmion in flat space resembles ``half'' a Skyrmion on a
$3$-sphere with optimal radius.  
This fits well with the observation in section
\ref{NumericalResults} that the parameter $R_{0,min}$ 
which is a measure of the size of
the Skyrmion agrees well with the optimal length $L_{opt.}^{numeric}$.

This line of thought can be carried even further. 
Skyrmions on the $3$-sphere might be a reasonable model
for nuclei, once they are quantized. 
The main advantage of this model is that on the $3$-sphere
one-loop corrections are expected to play a far less important 
role than in flat space. 
In particular, if the Skyrmion lives on the 
$3$-sphere with the optimal radius we expect these corrections
to be so small that the predictions of the Skyrme model could be
compared with experiment.
The main motivation for this claim are the promising results in
\cite{Hong-98} for the $B=1$ Skyrmion on the $3$-sphere, and the fact
that they predict a different value for the Skyrme coupling which is
in agreement with one-loop calculations performed in \cite{Meier-97}
and the Skyrme coupling therein.

\section*{Acknowledgements}

The author is grateful to N S  Manton for suggesting the
problem, and also wants to thank the participants of the 
DAMTP soliton seminar for fruitful discussions. 
The author thanks PPARC for a research studentship and the
Studienstiftung des deutschen Volkes for a PhD scholarship.

\appendix
\section*{Appendix}
\setcounter{section}{1}
\subsection{${\cal I}$ for $N = 1,\dots,9$ and $17$}

Table \ref{tableA1} shows the minimal value of the integral ${\cal I}$
in equation (\ref{calI}) as a
function of $N_R$, as has been calculated in \cite{Houghton-98}. 
``APPROX'' is the energy of the rational map ansatz for flat space
Skyrmions where the corresponding 
variational equation for the shape function has been solved 
numerically. The column ``TRUE'' shows the value of the energy
obtained by a numerical solution of the full set of partial
differential equations. 
The next column ``SYM'' shows the symmetries of the Skyrmions. The
numerical solutions and the rational map ansatz give the same
symmetry. Note that the respective values for $B=9$ in table
\ref{tableA1} are taken from \cite{Battye-00}. 
There is evidence that for $B=9$ the symmetry
and the other respective values in the table are different from the
ones cited in \cite{Houghton-98}. 

The last two columns are the optimal
length $L_{opt.}$ and the optimal energy $E_{opt.}^{numerical}$ which
are displayed in figure \ref{fig7} and \ref{fig6}, respectively.
As described in chapter \ref{NumericalResults}, 
$E_{opt.}^{numerical}$ has been obtained by using the rational map of
table \ref{tableA1} and calculating the shape function numerically. 

Following \cite{Houghton-98} 
we also include the values for $B = 17$. In this case the solution in
flat space has marginally lower energy than $E_{opt.}^{numeric}$. 
However, we still expect that the rational map ansatz is a good
approximation even in this situation.

\begin{table}[!h]
\begin{center}
\begin{tabular}{|c|c|c|c|c|c|c|}
\hline
 & & & & & & \\
$N_R$ & ${\cal I}$ & APPROX & TRUE & SYM & $L_{opt.}$ & $E_{opt.}^{numeric}$ \\
 & & & & & & \\
\hline
1 & 1.00  & 1.232 & 1.232 & $O(3)$ & 1.000 & 1.000 \\
2 & 5.81  & 1.208 & 1.171 & $O(2) \times {\mathbb Z}_2$ & 1.393 & 1.078 \\
3 & 13.58 & 1.184 & 1.143 & $T_d$ & 1.669 & 1.093 \\
4 & 20.65 & 1.137 & 1.116 & $O_h$ & 1.829 & 1.065 \\
5 & 35.75 & 1.147 & 1.116 & $D_{2d}$ & 2.074 & 1.089 \\
6 & 50.76 & 1.137 & 1.109 & $D_{4d}$ & 2.250 & 1.088 \\
7 & 60.87 & 1.107 & 1.099 & $Y_h$ & 2.346 & 1.064 \\
8 & 85.63 & 1.118 & 1.100 & $D_{6d}$ & 2.544 & 1.080 \\
\hline
9 & 109.3 & 1.098 & 1.093 & $ D_{4d} $ & 2.696 & 1.083 \\
\hline
17 & 367.41 & 1.092 & 1.073 & $Y_h$ & 3.614 & 1.076 \\
\hline
\end{tabular}
\end{center}
\caption{Values of ${\cal I}$, the energies in flat space (APPROX,
TRUE), the symmetry (SYM), and the optimal length $L_{opt.}$ and
energy $E_{opt.}^{numeric}$ on $S^3$.
All values are given for baryon number $B=N_R$ for
$B=1,\dots,9$ and $B=17$.\label{tableA1}} 
\end{table}

\subsection{The Rational Map Bound for Different Splittings $B=N_f N_R$}

In table \ref{tableA2} we compare the rational map bound for different
$N_f$. 
As the rational map bound is just the topological bound for
$N_R=1$, we have omitted $N_R=1$ from table \ref{tableA2}.

\begin{table}[!h]
\begin{center}
\label{Bogbounds}
\begin{tabular}{|c| |c|c| |c|c|c| |c|c|c| |c|c|}
\hline
 
$(N_f,N_R)$ & (1,4) & (2,2) & (1,6) & (2,3) & (3,2) & (1,8) & (2,4)
& (4,2) & (1,9) & (3,3)
\\
\hline
$E_{Bound}$  & 4.181 & 4.274 & 6.375 & 6.457 & 6.410 & 8.418 & 8.363
& 8.547 & 9.541 & 9.685
\\
\hline
\end{tabular}
\end{center}
\caption{The value of the generalized Faddeev-Bogomolny bound (\ref{Bograt})
for various values of $N_f$ and $N_R$.\label{tableA2}}  
\end{table}

Table \ref{tableA2} can be used to estimate which combination of $N_f$ and
$N_R$ might be more stable for a given baryon number $B$. 
It turns out that for $B=8$ the bound is lower for $N_f=2$
than for $N_f=1$. It is also probable that the $(N_f=3,N_R=2)$
Skyrmion has lower energy than the $(N_f=2,N_R=3)$ Skyrmion. 

\renewcommand{\baselinestretch}{.95}


\begin{thebibliography}{abcdefgh}
{\small
\bibitem{Skyrme-61} T H R Skyrme 1961
A nonlinear field theory
{\it Proc. Roy. Soc.} {\bf A260} 127.
\bibitem{Adkins-83} G Adkins, C Nappi and E Witten 1983 
Static properties of nucleons in the Skyrme model
{\it Nucl. Phys.} {\bf B228} 552.
\bibitem{Meier-97} F Meier and H Walliser 1997 
Quantum corrections to baryon properties in chiral soliton
models
{\it Phys. Rep.} {\bf 289} 383.
\bibitem{Leese-95} R A  Leese, N S  Manton and B J  Schroers 1995
Attractive channel Skyrmions and the deuteron
{\it Nucl. Phys.} {\bf B442} 228.
\bibitem{Irwin-00} P  Irwin 2000 
Zero mode quantization of multi-Skyrmions
{\it Phys. Rev.} {\bf D61} 114024.
\bibitem{Sutcliffe-97} R A  Battye and P M  Sutcliffe 1997
Symmetric Skyrmions
{\it Phys. Rev. Lett.} {\bf 79} 363.
\bibitem{Houghton-98} C J  Houghton, N S  Manton and P M  Sutcliffe
1998   
Rational maps, monopoles and Skyrmions 
{\it Nucl. Phys.} {\bf B510} 507.
\bibitem{Manton-87} N S  Manton 1987 
Geometry of Skyrmions
{\it Commun. Math. Phys.} {\bf 111} 469.
\bibitem{Castillejo-89} L Castillejo, P S J  Jones, A D  Jackson,
J J M  Verbaarshot, A Jackson 1989 
Dense Skyrmion systems
{\it Nucl. Phys.} {\bf A501} 801.
\bibitem{Kugler-88}  M  Kugler, S  Shtrikman 1988 
A new Skyrmion crystal
{\it Phys.Lett.} {\bf B208} 491.  
\bibitem{Kugler-89}  M  Kugler, S Shtrikman 1989
Skyrmion crystals and their symmetries
{\it Phys. Rev.} {\bf D40} 3421.  
\bibitem{Jackson-89} A D  Jackson, N S  Manton and
A  Wirzba 1989 
New Skyrmion solutions on a three sphere
{\it Nucl. Phys.} {\bf A495} 499.
\bibitem{Houghton-00} C J  Houghton and S Krusch 2001 Folding in the
Skyrme model {\it J. Math. Phys.} {\bf 42} 4079.
\bibitem{Eells-64} J Eells and J J  Sampson 1964
Harmonic mappings of Riemannian manifolds 
{\it Am. J. Math.} {\bf 86} 109.
\bibitem{Faddeev-76} L D  Faddeev 1976 
Some comments on the many dimensional solitons
{\it Lett. Math. Phys.} {\bf 1}  289.
\bibitem{Manton-Ruback} N S  Manton and P J Ruback 1986 
Skyrmions in flat space and curved space
{\it Phys. Lett.} {\bf B181} (1986) 137. 
\bibitem{Donaldson-84} S K Donaldson 1984
Nahm's equations and the classification of monopoles
{\it Commun. Math. Phys.} {\bf 96} 387.
\bibitem{Jarvis-96} S Jarvis 2000 A rational map for Euclidean
monopoles via radial scattering {\it J. Reine Angew. Math} {\bf 524} 17.
\bibitem{Loss-87} M Loss 1987 
The Skyrme model on Riemannian manifolds
{\it Lett. Math. Phys.} {\bf 14} 149.
\bibitem{Hong-98} S-T Hong 1998
The static properties of hypersphere Skyrmions
{\it Phys. Lett.} {\bf B417} 211.
\bibitem{Battye-00} R A  Battye and P M  Sutcliffe 2000 Skyrmions,
fullerenes and rational maps {\it Rev. Math. Phys.} {\bf 14} 29.
\label{lastref}
}
\end{thebibliography}
\end{document}